\documentclass[12pt]{article}
\RequirePackage{setspace}
    \AtBeginDocument{\setlength\abovedisplayskip{3pt}}
    \AtBeginDocument{\setlength\belowdisplayskip{3pt}}
\usepackage{amsfonts}
\usepackage{bbm}
\usepackage{graphicx}
\usepackage{sectsty}
\sectionfont{\normalsize}
\usepackage{appendix}
\usepackage{float}
\usepackage{url}
\usepackage{caption}
\usepackage[round]{natbib}
\begin{document}
\title{\Large The unreasonable success of quantum probability II:\\
\large Quantum measurements as universal measurements}
\author{\normalsize Diederik Aerts \\ 
        \small\itshape \vspace{-0.1 cm}
        Center Leo Apostel for Interdisciplinary Studies and Department  \\
        \small\itshape \vspace{-0.1 cm}
        of Mathematics, Brussels Free University, Brussels, Belgium \\ 
         \small \vspace{-0.1 cm}
        email: \url{diraerts@vub.ac.be} \vspace{0.2 cm} \\ 
             \normalsize   Massimiliano Sassoli de Bianchi  \\ 
       \vspace{-0.1 cm} \small\itshape
        Laboratorio di Autoricerca di Base,
         Lugano, Switzerland \\
        \small \vspace{-0.1 cm}
        email: \url{autoricerca@gmail.com} \vspace{0.2 cm}
        }
\date{}
\maketitle
\vspace{-1cm}
\begin{abstract}
\noindent 
In the first part of this two-part article \citep{AertsSassolideBianchi2014}, we have introduced and analyzed a multidimensional model, called the \emph{general tension-reduction} (GTR) model, able to describe general quantum-like measurements with an arbitrary number of outcomes, and we have used it as a general theoretical framework to study the most general possible condition of lack of knowledge in a measurement, so defining what we have called a \emph{universal measurement}. In this second part, we present the formal proof that universal measurements, which are averages over all possible forms of fluctuations,  produce the same probabilities as measurements characterized by \emph{uniform} fluctuations on the measurement situation. Since quantum probabilities can be shown to arise from the presence of such uniform fluctuations, we have proven that they can be interpreted as the probabilities of a first-order non-classical theory, describing situations in which the experimenter lacks complete knowledge about the nature of the interaction between the measuring apparatus and the entity under investigation. This same explanation can be applied -- mutatis mutandis -- to the case of cognitive measurements, made by human subjects on conceptual entities, or in decision processes, although it is not necessarily the case that the structure of the set of states would be in this case strictly Hilbertian. We also show that universal measurements correspond to maximally \emph{robust}  descriptions of indeterministic reproducible experiments, and since quantum measurements can also be shown to be maximally robust, this adds plausibility to their interpretation as universal measurements, and provides a further element of explanation for the great success of the quantum statistics in the description of a large class of phenomena.
\end{abstract}

\vspace{-0.1cm}

\vspace{-0.4cm}
\section{Introduction}
\label{Introduction}
\vspace{-0.4cm}

This is the second part of a two-part article, aimed at the investigation of a very general class of measurements, relevant to experiments performed both in cognitive science and physics. Although the present article is written in a self-consistent way, and therefore can be read independently of the content of its first part \citep{AertsSassolideBianchi2014}, the lecture of the latter is  recommended. We also refer the reader to the first article for a general discussion of the relevance of quantum structures in the macro world, and in particular in the field of psychology and cognitive science, referred to meanwhile commonly as `quantum cognition' \citep{AertsAerts1995, AertsBroekaertGaboraSozzo2013, AertsGabora2005a, AertsGabora2005b, AertsGaboraSozzo2013, Blutner2009, Blutner2013, BruzaetBusemeyerGabora2009, Bruzaetal2007, Bruzaetal2008a, Bruzaetal2008, Bruzaetal2009, Bruzaetal2009b, BusemeyerBruza2012, Busemeyeretal2011, Busemeyeretal2006, Franco2009, HavenKhrennikov2013, GaboraAerts2002, Khrennikov2010, KhrennikovHaven2009, PothosBusemeyer2009, VanRijsbergen2004, Wangetal2013, YukalovSornette2010}.

In \citet{AertsSassolideBianchi2014}, we have initially introduced and analyzed a model, that we have called the \emph{uniform tension-reduction} (UTR) model, allowing to represent probabilities associated with all possible one-measurement situations, and we have used it to explain the emergence of quantum probabilities (the Born rule) as \emph{uniform} fluctuations on the measurement situation. The particularity of the UTR-model is to exploit the geometry of simplexes to represent the states both of the measured and measuring entities, in a way that the outcome probabilities can be derived as the (uniform) Lebesgue measure of suitably defined convex subregions of the simplex under consideration. 

In \citet{AertsSassolideBianchi2014}, we have also shown that, although  the UTR-model is an abstract construct, it admits physical realizations, and we have proposed a very simple and evocative one, using a material point particle which is acted upon by special elastic membranes, which by breaking and collapsing are able to produce  the different possible outcomes. This easy to visualize mechanical realization allowed us to gain considerable insight into the hidden structure of a measurement process, be it that associated with a cognitive measurement, on a conceptual entity (or in the ambit of a decision process), or a physics measurement, on a microscopic entity. For instance, thanks to it, it becomes possible to visualize the structural difference between measurements performed on entangled entities (combinations of concepts, of elementary ``particles,'' etc.), as opposed to measurements performed on separate entities, in terms of a dimensional change in the measurement context, whose increased level of potentiality requires higher dimensional membranes to be described. 

Always in the first part of this two-part article, we have considered a further generalization of the UTR-model, by considering conditions of lack of knowledge which can be associated  with \emph{non-uniform} fluctuations, in what we have called the \emph{general tension-reduction} (GTR) model. In this more general theoretical framework, we have introduced and motivated a notion of \emph{universal measurement}, which describes the most general possible condition of lack of knowledge in a measurement, and we have pointed out that the uniform fluctuations characterizing quantum measurements can also be understood as an average over all possible forms of non-uniform fluctuations, which can in principle be actualized in a measurement context. 

The main purpose of this second part of the article is to present the formal proof of the theorem establishing the correspondence between universal measurements and uniform measurements. Since the quantum mechanical Born rule can be described in terms of uniform fluctuations, it follows from this result that quantum measurements can be interpreted as universal measurements, i.e., as measurements describing situations which are uniform mixtures of different possible measurement situations, thus defining a condition of lack of knowledge not only about the interaction which is each time actualized, between the measured entity and the measured system, but also about the way such interaction is chosen, among all the possible ways of choosing it. Of course, this doesn't mean that quantum measurements, performed on microscopic entities like electrons, neutrons, protons, etc., would  necessarily be universal measurements. For the time being, this remains an open question. What we know, however, is that the huge average involved in a universal measurement is perfectly compatible with this interpretation. Hence, we have proven that the hypothesis that a quantum measurement is a universal measurement can be true, and hence, since the way in which probability arises and is intrinsic in a universal measurement can be understood intuitively and completely, in case this hypothesis is true, it entails a deep explanation of the nature of a quantum measurement. 

For cognitive measurements the situation is different. Indeed, a typical experiment in cognitive science usually involves different human subjects who, because of the specificities and uniqueness of their mind structures, will necessarily have different ways of choosing the possible answers to the questions that are addressed to them, or will have different ways of deciding when facing situations requiring a decision process. Therefore, their actions will be in principle characterized by  different sets of probabilities for the different outcomes. In other terms, if we consider the ensemble of the participants in a given experiment, we can say that the overall statistics of outcomes they deliver is the result of an average over different kinds of measurements. This not only because each subject would choose/decide according to different criteria, but also because even a same subject can choose according to different criteria in two different moments. This means  that a model where the probabilities associated with the different possible outcomes, obtained from the answers collected from the different subjects, are considered as averages over the different probabilities associated with the different measurements performed by each of them, is a model that corresponds well to the situation we imagine to occur.

So, in cognitive science, if we want to be adherent to what can readily be supposed to actually happen during an experiment, modelizing a measurement situation in the form of a meta-measurement, i.e., in a way that the statistics of outcomes results from a process of randomization over different individual measurements,  is a plausible approach. This is precisely what the GTR-model allows to do, describing the different possible measurements by means of different probability densities $\rho$, which are subsequently averaged out in what we have called a universal measurement. 

Considering the above, it is clear that the result of the equivalence between universal measurements and uniform measurements acquires a different meaning in quantum physics and cognitive science. In quantum physics it has more the value of an explanation of the nature of quantum probabilities. It suggests that the level of potentiality inherent in a microphysical quantum process is possibly much deeper than it was initially hypothesized in the so-called \emph{hidden measurement approach} \citep{Aerts1986, Aerts1998, Aerts1999b, Coecke1995, SassolideBianchi2013}, as it would concern not only the process of actualization of single measurement interactions, but of entire ways of choosing these interactions. 

For the cognitive ambit, it predicts the existence of a `first order theory,' which may be equal to the orthodox quantum theory, but can also be different from it, but can in any case be formulated using a uniform probability density $\rho_u$, in what we have called the UTR-model. This possible difference between orthodox quantum mechanics and the `first order theory' describing cognitive measurements, would manifest at the level of the structure of the set of states of the conceptual entities,\footnote{We recall that it is possible to formalize a concept as an \emph{entity in a specific state}, and a context as a ``surrounding'' which is able to produce a change (either deterministic or indeterministic) of such state  \citep{AertsGabora2005a, GaboraAerts2002}.} which is not necessarily Hilbertian. If it will turn out to be Hilbertian, as apparently is the case for the microphysical entities,\footnote{The majority of physicists would generally agree that orthodox quantum theory does conveniently describe all possible measurements on entities of the micro world, i.e., that the Hilbert space model, equipped with the Born rule, is sufficient to capture the essence of the behavior of microscopic entities, when acted upon in the different measurement contexts. But maybe this is something we should not take as fully granted. This because, as was noted by one of us in the eighties, the theoretical construct of orthodox quantum physics presents some severe structural shortcomings \citep{Aerts1982, Aerts1986, Aerts1999a, AertsDurt1994, Aertsetal1997a, Aertsetal1999}. For instance, it cannot describe the situation where entities can become fully separated in experimental terms. However, the possibility of pointing out, also at the microscopic level, the existence of separated entities, cannot be excluded. As an example, consider the famous coincidence measurements conducted by Alain Aspect on entangled photons \citep{Aspect1999}. These measurements are usually adjusted in a way as to only select pairs of entities that, as they fly apart, remain connected, thus producing a violation of Bell's inequalities. But experiments could also be adjusted in a way that possible coincident but disconnected pairs could also be detected, not necessarily giving rise to correlations. This possibility would describe regimes which are non-quantum, but quantum-like, and therefore not describable by the Born rule. The problem is that this possibility is usually not investigated, as experimenters will generally not consider adjustments of their measurements so as to allow for the possibility that entangled entities could also disentangle during their fly, as this would in general be interpreted as a badly performed experiment. However, these ``wrong'' adjustments, associated with ``badly performed experiments,'' could also in principle be interpreted as adjustments that favor the selection of measurements which, if taken into consideration, would lead to a statistics of outcomes different from that of Born. In other terms, we cannot totally exclude that in some of our quantum measurements we are maybe filtering out some of the outcomes (unduly considering them as outliers), and this could be the reason why we are not yet able to observe possible deviations from the Born statistics.} then cognitive measurements, understood as universal measurements, will also turn out to be equivalent to quantum measurements, in the sense that not only the probabilities of single measurements, but also of sequential and conditional measurements, will produce the same values as those predicted by the Born rule. On the other hand, if it will  turn out not to be Hilbertian, considering however the effectiveness of so many quantum models of cognition and decision (and the proof of the equivalence between universal and uniform measurements presented in this article), we might nevertheless expect  the theory behind such models of cognition and decision to  be quite close to quantum theory. Also because we already know that cognitive measurements do reveal the presence of typical quantum effects of interference, contextuality, entanglement and emergence \citep{Aerts2009, Aertsetal2000, AertsGabora2005a, AertsGabora2005b, AertsSozzo2011, GaboraAerts2002}, when concepts in different possible states are combined and data of experiments with human subjects on such combinations of concepts are modeled. But this of course does not mean that all these effects necessarily originate from a strict Hilbertian structure for the set of states. For instance, as emphasized in \citet{AertsSozzo2012a, AertsSozzo2012b}, one doesn't need linearity to model entanglement. 

It is worth emphasizing that to be able to identify what is the probability model which is behind cognitive measurements (assuming that a single explanatory model can consistently describe all possible data), one needs a sufficiently general theoretical framework to integrate and analyze not only the existing data, but also, and more importantly, the data that will become available in the future. Here it is important to distinguish `ad hoc mathematical models,' constructed only with the purpose of fitting  data, from more `fundamental models,' which try to identify, in a more stringent way, the basic structure which is really behind the collected data, as well as its possible significance. In physics, for instance, one usually speaks of `phenomenological models,' which just organize mathematically the results of the observed phenomena without paying too much attention to their possible significance, as opposed to more deep `explanatory models,' which instead try to understand the observed phenomena at a more fundamental level, possibly deriving them from first principles.

The GTR-model, and the proposed notion of \emph{universal measurement}, is precisely a first step towards the construction of a more fundamental and explanatory model, able not only to phenomenologically  account for the different data, but also to possibly explain their origin. The theorem that we prove in the present article, can be considered as an attempt of constructing cognitive models starting also from first principles. Here the first principle we advocate  is that a cognitive measurement is essentially an averaged measurement (described in the GTR-model by the different $\rho$ characterizing the different ways of choosing the outcomes), and that if the average involves a sufficiently large set of data, then a `first order theory' will emerge, which is described by a simpler structure in which probabilities are obtained in terms of the uniform Lebesgue measure (the UTR-model). 

It is worth 
%Massimiliano: emphasizing 
mentioning
that the GTR-model, contrary to the classical Kolmogorovian model, has the advantage of also allowing for a representation of the states of the entity under consideration (be it conceptual, or physical), as points in a simplex. Also the Hilbert-model, of course, allows for a representation of the states, but it imposes them, from the start, a very specific (Hilbertian) structure. In cognitive science, however, as we previously mentioned, we haven't yet determined what is the structure of the set of states. To determine that structure, one needs to go beyond the analysis of single-measurement situations, and collect data from which also conditional and/or sequential probabilities can be deduced. However, these data will not be analyzable within the too limited structure of the  Hilbert-model, but will require the more ample framework of the UTR-model (which is not necessarily linear) or of the GTR-model (which is not necessarily uniform). 

Let us also mention another advantage of the GTR-model: the fact that, similarly to the Hilbert-model, it contains a procedure for forming joint (possibly entangled) entities, by simply multiplying the dimension of the real spaces describing the single entities \citep{AertsSassolideBianchi2014}, but, different from the Hilbert-model, it does so without the hypothesis of linearity. This possibility is of course absent in classical Kolmogorovian models, which by the way are also incapable of describing measurement processes which change the state of the measured entity (which is the typical situation both in microphysics and cognitive sciences). All the above advantages will certainly prove to be instrumental in future works, aimed at clarifying the (possibly non-Hilbertian) structure of the set of states of human conceptual entities. 

In addition to the description of the GTR-model, and the proof of the equivalence between universal and uniform measurements, we shall also discuss in this second part of the article the important notion of `\emph{robustness} of a measurement,' showing that universal measurements correspond to maximally \emph{robust}  indeterministic reproducible experiments. This  characteristic being also shared by quantum measurements, as evidenced in the recent analysis of \citet{DeRaedtetal2013}, the result not only increases the plausibility that quantum measurements are also universal measurements, but provides a further explanation of their success in describing different fields of investigations.

The work is organized as follows. In Sec.~\ref{Quantum Probabilities for a Single Observable}, we present some basic elements of the quantum formalism, to define notations and to allow to easily establish, in the subsequent sections, the correspondence between  quantum measurements and the measurement described in the GTR-model, when the probability density $\rho$ is uniform. In Sec.~\ref{Therhomodel}, we introduce the GTR-model and recall that in the uniform case it becomes isomorphic to the Hilbert-model of quantum mechanics, when the states describing the entity under investigation come from a Hilbert space. Then, in Sec.~\ref{theorem}, we define the fundamental notion of \emph{universal measurement} and enunciate the theorem establishing the equivalence between  universal  and uniform measurements. 

To prove the theorem, we proceed with the following steps. In Sec.~\ref{limit of cellular structures}, we discretize the GTR-model and show that one can always consider limits of cellular probability densities to approximate, with arbitrary precision, the probabilities obtained from arbitrary, non-cellular, probability densities. This will allow us to consider, in Sec.~\ref{Averaging over finite cellular structures}, the average over all possible kinds of measurements, focusing our analysis on discretized structures, proving in this way the theorem. 

%Massimiliano: In Sec.~\ref{nonhilbert}, we  analyze in some detail the ampler structural richness of the $\rho$-model, by investigating, in the two-outcome case, and in the ambit of the so-called \emph{sphere-model}, the probabilities associated with sequential measurements, showing that they cannot in general be fitted into a Hilbertian or Kolmogorovian structure. 
Finally, in Sec.~\ref{Robustness}, we study the notion of \emph{robustness}, and show that universal measurements are maximally robust, and in Sec.~\ref{Conclusion} we offer some conclusive remarks.

\vspace{-0.4cm}
\section{Quantum Probabilities for a Single Observable\label{quantumprobability}}
\label{Quantum Probabilities for a Single Observable}
\vspace{-0.4cm}

In this section we present the basic formalism of quantum mechanics, in relation to the measurement of a finite dimensional observable. Not to unnecessarily complicate the discussion, we shall limit ourselves, throughout the article, to the situation of a non-degenerate measurement, associated with a non-degenerate observable. All the results we shall obtain  can be easily generalized to the degenerate case, proceeding as shown in \citep{AertsSassolideBianchi2014}.

In orthodox quantum theory, the state of an entity (for a physicist it can be a microscopic entity, like an neutron, for a cognitive scientist, a concept, or a situation apt for a decision process)  is described by a unit vector of a vector space over the field ${\mathbb C}$ of complex numbers -- the so-called Hilbert space ${\mathcal H}$ -- equipped with a (sesquilinear) inner product $\langle \cdot |  \cdot \rangle$, which maps two vectors $|\phi\rangle$, $|\psi\rangle$ to a complex number $\langle \phi|\psi\rangle$, and consequently with a norm $\| |\psi\rangle \| \equiv \sqrt{\langle \psi|\psi\rangle}$, which assigns a positive length to each vector. In this paper we only consider  Hilbert spaces having a finite number of dimensions, and we  denote  $\mathcal{H}_{N}$ a Hilbert space whose vectors are $N$-dimensional.

An observable is a measurable quantity of the entity under consideration, and in quantum theory is represented by a self-adjoint operator $A$, acting on vectors of the Hilbert space, i.e., $A: |\psi\rangle \rightarrow A|\psi\rangle$. In our case, being the Hilbert space $N$-dimensional,  $A$  can be entirely described by means of its $N$ eigenvectors $|a_i\rangle$ and the associated (real) eigenvalues $a_i$, obeying the eigenvalue relations $A|a_i\rangle = a_i |a_i\rangle$, for all $i\in \{1,\dots,N\}\equiv I_{N}$. If the eigenvectors have been duly normalized, so that in addition to the orthogonality relation $\langle a_i|a_j\rangle = \delta_{ij}$, $i,j\in I_{N}$, they also obey the completeness relation $\sum_{i\in I_{N}} |a_i\rangle\langle a_i| = \mathbb{I}$, where $\mathbb{I}$ denotes the unit operator, they can be used to construct the orthogonal projections $P_i\equiv |a_i\rangle\langle a_i|$, $i\in I_{N}$, obeying $\sum_{i\in I_{N}} P_i =\mathbb{I}$, $P_iP_j =P_i\delta_{ij}$, $i,j\in I_{N}$, which in turn can be used to write the observable $A$ as the (spectral) sum:
\begin{eqnarray}
\label{observable-hilbert}
A = \mathbb{I} A = \left[\sum_{i\in I_{N}} P_i\right]A=\sum_{i\in I_{N}} a_i P_i.
\end{eqnarray}

Similarly, if $|\psi\rangle\in \mathcal{H}_{N}$, $\|\psi\|^2=\langle \psi|\psi\rangle=1$, is a normalized vector describing the state of the entity, it can be written as the sum: 
\begin{eqnarray}
\label{state-hilbert}
|\psi\rangle = \mathbb{I} |\psi\rangle = \left[\sum_{i\in I_{N}} P_i\right]|\psi\rangle =\sum_{i\in I_{N}} |a_i\rangle\langle a_i|\psi\rangle = \sum_{i\in I_{N}} \sqrt{x_i}e^{i\alpha_i} |a_i\rangle,
\end{eqnarray}
where for the last equality we have written the complex numbers $\langle a_i|\psi\rangle$ in the polar form $\langle a_i|\psi\rangle = \sqrt{x_i}e^{i\alpha_i}$. Clearly, being $|\psi\rangle$ normalized to $1$, the positive real numbers $x_i$ must obey:
\begin{eqnarray}
\label{sumofthexi}
\sum_{i\in I_{N}}x_i = 1.
\end{eqnarray}

When we measure an observable $A$ in a practical experiment (a physicist does so by letting the microscopic entity interact with a macroscopic measuring apparatus, a psychologist by letting a human concept interact with a human mind, according to a certain protocol, if concepts are studied, or by collecting the decision results, if situations lending themselves to human decisions are studied), we can obtain one of the $N$  eigenvalues $a_i$, $i\in I_{N}$, and if  these $N$ eigenvalues are all different, we say that the spectrum of $A$ is \emph{non-degenerate}. Consequently, the measurement has $N$ distinguishable possible outcomes. 

In general terms, the measurement of an observable $A$ is a process during which the state of the entity undergoes an abrupt transition -- called ``collapse'' in the quantum jargon -- passing from the initial state $|\psi\rangle$ to a final state which is one of the eigenvectors $|a_i\rangle$ of $A$, associated with the eigenvalue $a_i$, $i\in I_{N}$. The process is non-deterministic, and we can only describe it in probabilistic terms, by means of a ``golden rule'' called the \emph{Born rule}, which states the following: the probability $P(|\psi\rangle\to|a_i\rangle)$ for the transition $|\psi\rangle\to|a_i\rangle$, is given by the  square of the length of the vector $P_i|\psi\rangle$, i.e., the square of the length  of the initial vector, once it has been projected onto the eigenspace of $A$ corresponding to the eigenvalue $a_i$. More explicitly: 
\begin{eqnarray}
\label{Born}
\lefteqn{P(|\psi\rangle\to|a_i\rangle) = \| P_i|\psi\rangle\|^2 = \langle\psi|P_i P_i|\psi\rangle}\nonumber\\
&= \langle\psi| P_i|\psi\rangle = \langle\psi|a_i\rangle\langle a_i|\psi\rangle =|\langle a_i|\psi\rangle|^2 = x_i,
\end{eqnarray}
for all $i\in I_{N}$. And of course, following (\ref{sumofthexi}), we have 
\begin{eqnarray}
\label{totalprob=1}
\sum_{i\in I_{N}} P(|\psi\rangle\to|a_i\rangle) = \sum_{i\in I_{N}}x_i = 1.
\end{eqnarray}

\vspace{-0.4cm}
\section{The GTR-model}
\label{Therhomodel}
\vspace{-0.4cm}

In this section we present the GTR-model, which is able to describe very general situations of measurement, characterized by an arbitrary (finite) number of outcomes. To facilitate the understanding of its logic, we present the model as an idealized mechanical model. This, however, must not mislead us: the GTR-model is an abstract construct, totally independent of its possible physical realizations. These very general measurement situations include those associated with classical, almost deterministic measurements, when the outcomes are fully predictable (if the state of the entity is known), but also, more generally,  those associated with quantum-like measurements, when the outcomes are most of the times genuinely unpredictable, but  nevertheless characterizable in probabilistic terms.  

The construction of the model draws its inspiration from a previous two-outcome model, called the \emph{sphere-model} (see  Sec.~\ref{Averaging over finite cellular structures}), which in turn is a generalization of the so-called $\epsilon$\emph{-model}, originally designed to study two-state systems, like spin-${1\over 2}$ quantum entities, and their generalizations \citep{Aerts1998, Aerts1999b, SassolideBianchi2013, AertsSassolideBianchi2014}. 

The GTR-model has many interesting features. It generalizes the sphere-model, allowing for the description of measurements having an arbitrary number of outcomes, and it offers a considerable insight into the internal working of quantum and quantum-like systems,  as it allows for a full visualization of what goes on during a measurement. The paradigm at the basis of its operation is that of the \emph{hidden-measurement approach} \citep{Aerts1986, Aerts1998, Aerts1999b, SassolideBianchi2013}, where the emergence of quantum structures is explained as the consequence of the presence of fluctuations in the measurement context. In this approach, the indeterminism inherent in quantum and quantum-like systems is due to the fact that a measurement is made of different possible \emph{pure measurements} (almost deterministic interactions), which can be actualized in an unpredictable way during the execution of the experiment. 

These pure measurements are hidden in the sense that  a physicist, when experimenting with microscopic entities, cannot distinguish them at the macroscopic level. Similarly, in cognitive experiments, they are hidden because they are actualized at the subconscious level, via ``non-logical'' intrapsychic processes, which cannot be discriminated at the conscious level. The great explanatory power of the GTR-model resides  precisely in the fact that these hidden aspects of a measurement process are  made fully manifest, so that one can really ``see'' what is going on, during its execution.

Another interesting aspect of the GTR-model is that it can describe all sorts of probabilistic models, generally non-Kolmogorovian and non-Hilbertian \citep{AertsSassolideBianchi2014}, and thanks to the very general theoretical framework it provides, it can be exploited to reveal an even more hidden aspect which is possibly at the basis of quantum probability, and which can explain its ``unreasonable'' success in the description of so many kinds of measurement situations, in different layers of reality. This hidden aspect is the fact that quantum measurements turn out to be interpretable as processes of randomization not only over deterministic -- pure measurement -- interactions, but also, and more generally, over all possible kinds of measurements. In other terms, as we will demonstrate in the next sections, the  GTR-model allows to explain quantum measurements as \emph{universal measurements}, i.e., as measurements expressing a much deeper level of actualization of potential elements of reality. 

\begin{figure}[!ht]
\centering
\includegraphics[scale =.65]{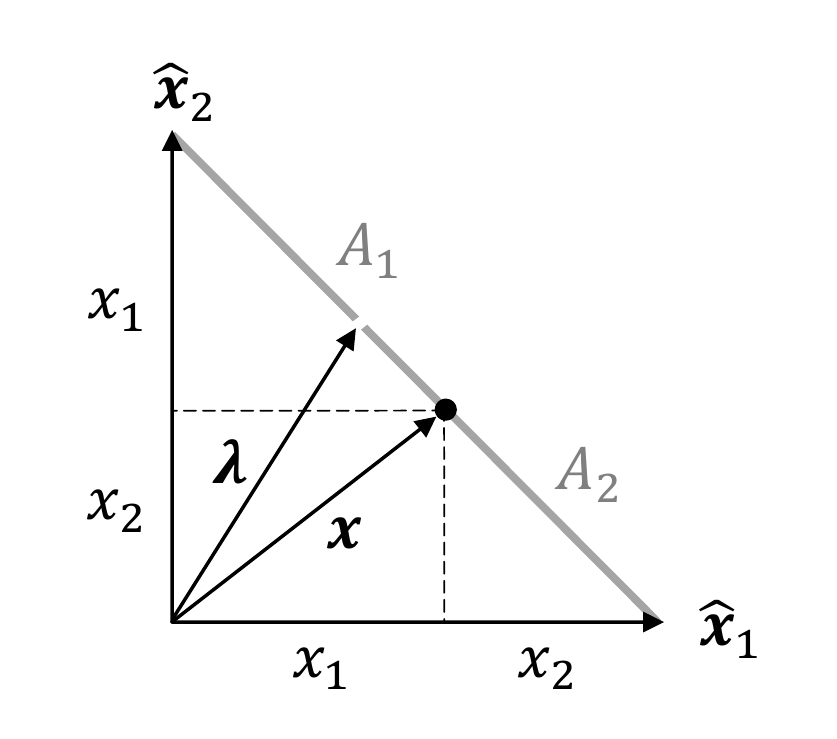}
\caption{A $1$-dimensional $\rho$-elastic structure attached to the two unit vectors $\hat{\bf{x}}_1$ and $\hat{\bf{x}}_2$, with the two  regions $A_1$ and $A_2$ generated by the presence of the particle in $\bf{x}$. The vector \mbox{\boldmath$\lambda$}, here in region $A_1$, indicates the point at which the elastic breaks. 
\label{1-dimensions}}
\end{figure}

To explain the functioning of the model, we proceed step by step, describing first the two-outcome situation ($N=2$), then the three-outcomes situation ($N=3$), and finally the general situation, with an arbitrary number of outcomes. As we said in the previous section, we shall limit our discussion to non-degenerate measurements, and refer to \citep{AertsSassolideBianchi2014} for the description of the degenerate situation.\footnote{In \citet{AertsSassolideBianchi2014} we only describe the degenerate situation in the special case of the UTR-model, i.e., in the special case where the probability density $\rho$ is uniform. However, the same description holds, mutatis mutandis, for the non-uniform case.}
\\
\vspace{-0.3cm}
\\
\emph{The $N=2$ case, with two outcomes}
\\
We consider a material point particle living in a Euclidean space ${\mathbb R}^{n}$, $n\geq 2$. Measurements, which will be denoted $e_{\{1\}\{2\}}^\rho$, can only have two outcomes.
The procedure to follow to perform $e_{\{1\}\{2\}}^\rho$ is the following. The experimenter takes a sticky breakable elastic band of the $\rho$-kind (what this means will be explained shortly), and stretches it over a $1$-dimensional simplex\footnote{A simplex is a generalization of the notion of a triangle. A 1-simplex is a line segment; a 2-simplex is an equilateral triangle; a 3-simplex is a tetrahedron; a 4-simplex is a pentachoron; and so on.} $S_1$, generated by two orthonormal vectors $\hat{\bf{x}}_1$ and $\hat{\bf{x}}_{2}$.  Once the $\rho$-elastic band is in place, the particle, by moving deterministically towards it (along a trajectory that is not important here to specify), sticks to it at a particular point ${\bf{x}}=x_1 \hat{\bf{x}}_1 + x_2 \hat{\bf{x}}_2$, $x_1 + x_2 = 1$, defining the state of the particle on the elastic (we represent Euclidean vectors in bold). When this happens, two disjoint regions $A_1$ and $A_2$ can be distinguished, which are respectively the region bounded by vectors $\hat{\bf{x}}_2$ and $\bf x$, and the region bounded by vectors $\bf x$ and $\hat{\bf{x}}_1$ (see Fig.~\ref{1-dimensions}).

\begin{figure}[!ht]
\centering
\includegraphics[scale =.55]{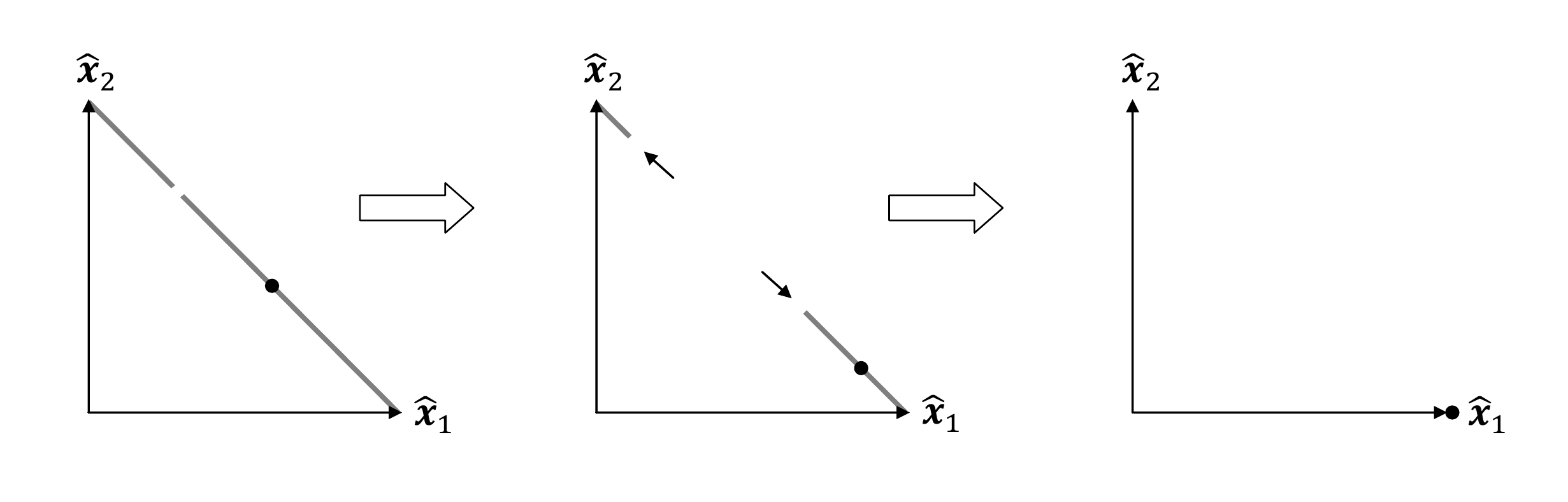}
\caption{The breaking of the $\rho$-elastic causes the particle to be drawn to  point $\hat{\bf{x}}_1$.  
\label{1-dimensionsbreaking}}
\end{figure}

Then, after some time, as it is made of a breakable material, the $\rho$-elastic inevitably breaks, at some a priori unpredictable point \mbox{\boldmath$\lambda$}. If $\mbox{\boldmath$\lambda$}\in A_1$, the band, by contracting,  draws the particle to point $\hat{\bf{x}}_1$ (the ``collapse process'' depicted in Fig.~\ref{1-dimensionsbreaking}), whereas if $\mbox{\boldmath$\lambda$}\in A_2$, it  draws the particle to point $\hat{\bf{x}}_2$.
We can observe that to each breaking point \mbox{\boldmath$\lambda$}, it corresponds a specific interaction between the particle and the elastic band, which deterministically draws the former  to its final state $\hat{\bf{x}}_1$, or $\hat{\bf{x}}_2$. In other terms, the measurement $e_{\{1\}\{2\}}^\rho$ is formed by a collection of hidden \emph{pure measurements}, i.e., potential, almost deterministic measurement interactions, only one of which is each time selected (actualized), when the elastic breaks. These pure measurements are \emph{almost} deterministic because if $\mbox{\boldmath$\lambda$} ={\bf x}$, then the interaction remains indeterminate, in the classical sense of a system in a condition of unstable equilibrium.  

To calculate the probabilities of the two possible outcomes, one needs to know what is the physical mechanism governing the breaking of the elastic structure. It is possible to imagine elastics that break in an infinite number of different ways, depending on their nature and the manner in which they were manufactured. In general terms, we can assume that each different elastic band can be characterized by a probability density $\rho: S_1 \to [0,\infty [$, describing the probabilities for the elastic of breaking in the different regions of $S_1$. Accordingly, the probability $P(A_i|\rho)$ for a $\rho$-elastic (i.e., an elastic characterized by the probability density $\rho$) to break in the region $A_i$, $i=1,2$, is given by the integral:   
\begin{eqnarray}
P(A_i|\rho)=\int_{A_i} \rho({\bf{y}})d{\bf{y}},
\end{eqnarray}
and of course:
\begin{eqnarray}
P(S_1|\rho)=P(A_1|\rho) + P(A_2|\rho) = \int_{S_1} \rho({\bf{y}})d{\bf{y}} = 1.
\end{eqnarray}
 
Now, since the particle is drawn to  $\hat{\bf{x}}_{i}$ when the elastic breaks in $A_i$, the probability $P({\bf{x}}\to \hat{\bf{x}}_{i}|\rho)$ for the transition ${\bf{x}}\to \hat{\bf{x}}_{i}$, is precisely the probability $P(A_i|\rho)$ for the elastic to break in $A_i$, so that we can write: 
\begin{eqnarray}
\label{probability-2-outcomes}
P({\bf{x}}\to \hat{\bf{x}}_{1}|\rho)=\int_{A_1} \rho({\bf{y}})d{\bf{y}}, \quad P({\bf{x}}\to \hat{\bf{x}}_{2}|\rho)=\int_{A_2} \rho({\bf{y}})d{\bf{y}}.
\end{eqnarray}
To obtain more explicit expressions, we observe that ${\bf{y}} = y_1 \hat{\bf{x}}_1 + y_2 \hat{\bf{x}}_2$, so that the first integral in (\ref{probability-2-outcomes}) can be written as the double integral: 
\begin{eqnarray}
\label{x1integralbis}
P({\bf{x}}\to \hat{\bf{x}}_{1}|\rho)= \int_{A_1}\rho(y_1,y_2)dy_1dy_2,
\end{eqnarray}
and since $y_1+y_2 = 1$, we can introduce the new variables:
\begin{eqnarray}
z_1&=&{y_1-y_2\over\sqrt{2}}, \quad z_2={y_1+y_2\over\sqrt{2}} = {1\over\sqrt{2}}\\
y_1&=&{z_1+ z_2\over\sqrt{2}} ={z_1\over\sqrt{2}} + {1\over2}, \quad y_2 ={z_2- z_1\over\sqrt{2}}= - {z_1\over\sqrt{2}}+{1\over2}, 
\end{eqnarray}
so that  the  double integral (\ref{x1integralbis}) transforms into the single integral:
\begin{eqnarray}
\label{x1integral-z}
P({\bf{x}}\to \hat{\bf{x}}_{1}|\rho) = \int_{-{1\over\sqrt{2}}}^{-{1\over\sqrt{2}}(1-2x_1)} \rho(z) dz,
\end{eqnarray}
where we have defined the one-dimensional probability density: $\rho(z)\equiv \rho({z\over\sqrt{2}} + {1\over2}, - {z\over\sqrt{2}}+{1\over2})$.

An important special case is that of a uniform probability density $\rho_u$, describing an elastic band made of a perfectly uniform material (so that all its segments have the same chance of breaking). Since $S_1$ has length $\|\hat{\bf{x}}_{2}-\hat{\bf{x}}_{1}\| = \sqrt{2}$, $\rho_u(z)={1\over\sqrt{2}}$, and (\ref{x1integral-z}) becomes: 
\begin{eqnarray}
\label{x1integral-z-u}
P({\bf{x}}\to \hat{\bf{x}}_{1}|\rho_u)= {1\over\sqrt{2}}\int_{-{1\over\sqrt{2}}}^{-{1\over\sqrt{2}}(1-2x_1)} dz = x_1.
\end{eqnarray}
In other terms, in accordance with the analysis of the UTR-model presented in \citet{AertsSassolideBianchi2014}, we obtain a result in agreement with the Born rule (\ref{Born}), i.e., the uniform measurement $e_{\{1\}\{2\}}^{\rho_u}$ is  isomorphic to the measurement of an observable $A$ in a two-dimensional complex Hilbert space ${\mathcal H}_2$, if we  represent the quantum state vector $|\psi\rangle = \sqrt{x_1}e^{i\alpha_1}|a_1\rangle +\sqrt{x_2}e^{i\alpha_2}|a_2\rangle \in {\mathcal H}_2$, whose components give the probabilities for the transitions $|\psi\rangle \to |a_1\rangle$ and $|\psi\rangle \to |a_2\rangle$ (once their square modulus is taken), and to that quantum state vector we associate a real vector ${\bf{x}} = x_1 \hat{\bf{x}}_1 + x_2 \hat{\bf{x}}_2$, in the 1-simplex $S_1$, whose components give these same probabilities for the corresponding transitions ${\bf{x}}\to \hat{\bf{x}}_{1}$ and ${\bf{x}}\to \hat{\bf{x}}_{2}$ (see Sec.~\ref{Quantum Probabilities for a Single Observable}).
\\
\vspace{-0.3cm}
\\
\emph{The $N=3$ case, with three outcomes}
\\
We consider now the slightly more elaborate situation consisting of three possible outcomes. The entity is always a material point particle,  living in a Euclidean space ${\mathbb R}^{n}$, $n\geq 3$. Different typologies of (non-trivial) measurements can be carried out in this case.  More precisely, we can distinguish four different typologies of measurements: $e_{\{1\}\{2\}\{3\}}^\rho$, $e_{\{1,2\}\{3\}}^\rho$, $e_{\{1,3\}\{2\}}^\rho$ and $e_{\{2,3\}\{1\}}^\rho$. We shall only consider the first one, which corresponds to the situation where all three outcomes can be distinguished by the experimenter (non-degenerate measurement), and refer the reader to \citet{AertsSassolideBianchi2014}, for the description of the other degenerate situations, where not all outcomes can be distinguished. 

\begin{figure}[!ht]
\centering
\includegraphics[scale =.7]{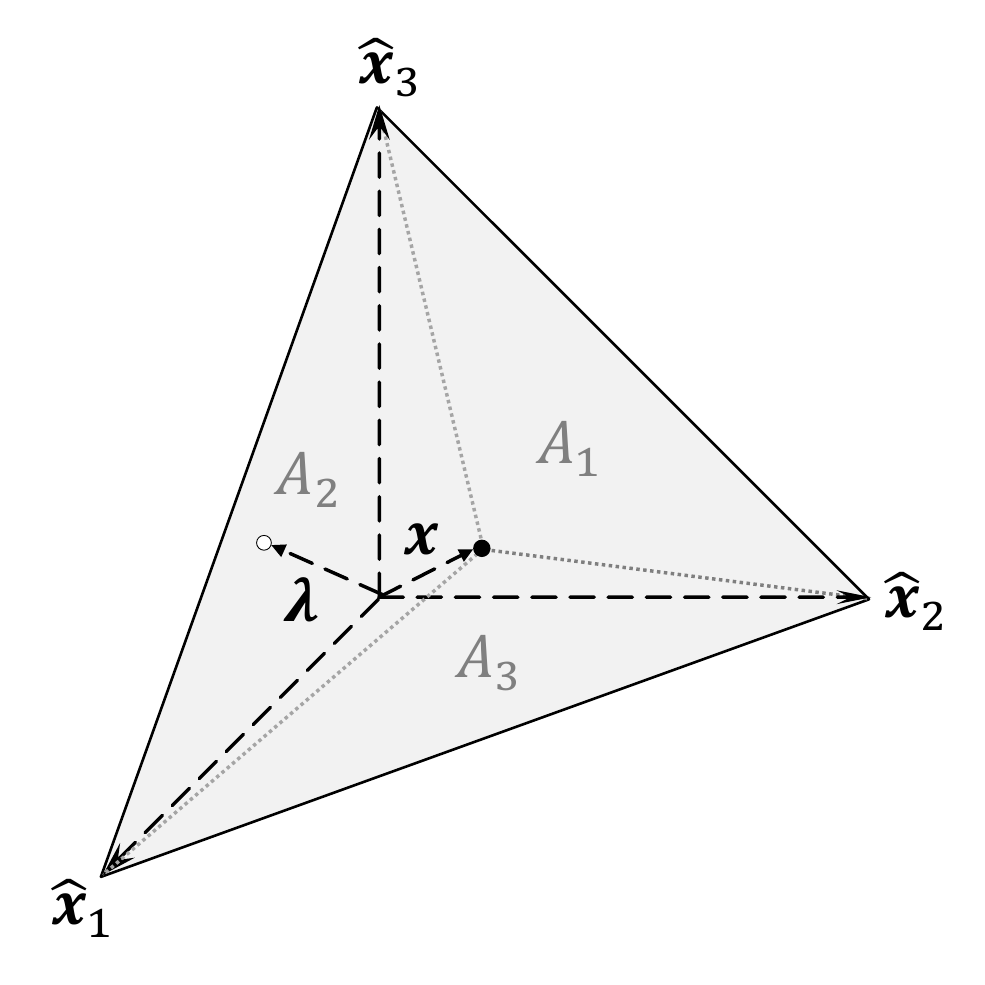}
\caption{A $2$-dimensional triangular $\rho$-membrane attached to  the three vertex vectors $\hat{\bf{x}}_1$, $\hat{\bf{x}}_2$ and $\hat{\bf{x}}_3$, with the three disjoint convex regions $A_1$, $A_2$, and $A_3$, generated by the presence of the particle in $\bf{x}$ (the ``tension lines'' of demarcation between the three regions correspond to the clear  dashed lines in the drawing). The vector \mbox{\boldmath$\lambda$}, here in region $A_2$, indicates the point where the elastic membrane breaks.
\label{triangolo}}
\end{figure}

The procedure to follow to perform $e_{\{1\}\{2\}\{3\}}^\rho$ is the following. The experimenter takes a sticky breakable elastic membrane of the $\rho$-kind, and stretches it over a $2$-dimensional simplex $S_2$ generated by  three orthonormal  vectors $\hat{\bf{x}}_1$, $\hat{\bf{x}}_2$ and $\hat{\bf{x}}_{3}$, attaching it to its three vertex points. Once the  membrane is in place, the particle, by moving deterministically towards it (along a trajectory that is not important here to specify), sticks to it at a particular point ${\bf{x}}=x_1 \hat{\bf{x}}_1 + x_2 \hat{\bf{x}}_2+ x_3 \hat{\bf{x}}_3$, with $x_1 + x_2 + x_3 = 1$, defining the state of the particle on the membrane. When this happens, three different disjoint convex regions $A_1$, $A_2$ and $A_3$ can be distinguished on the membrane's surface, delimited by  three ``tension lines'' which connect  $\bf{x}$ to the  vertex points of $S_2$ (see Fig.~\ref{triangolo}). 

\begin{figure}[!ht]
\centering
\includegraphics[scale =.45]{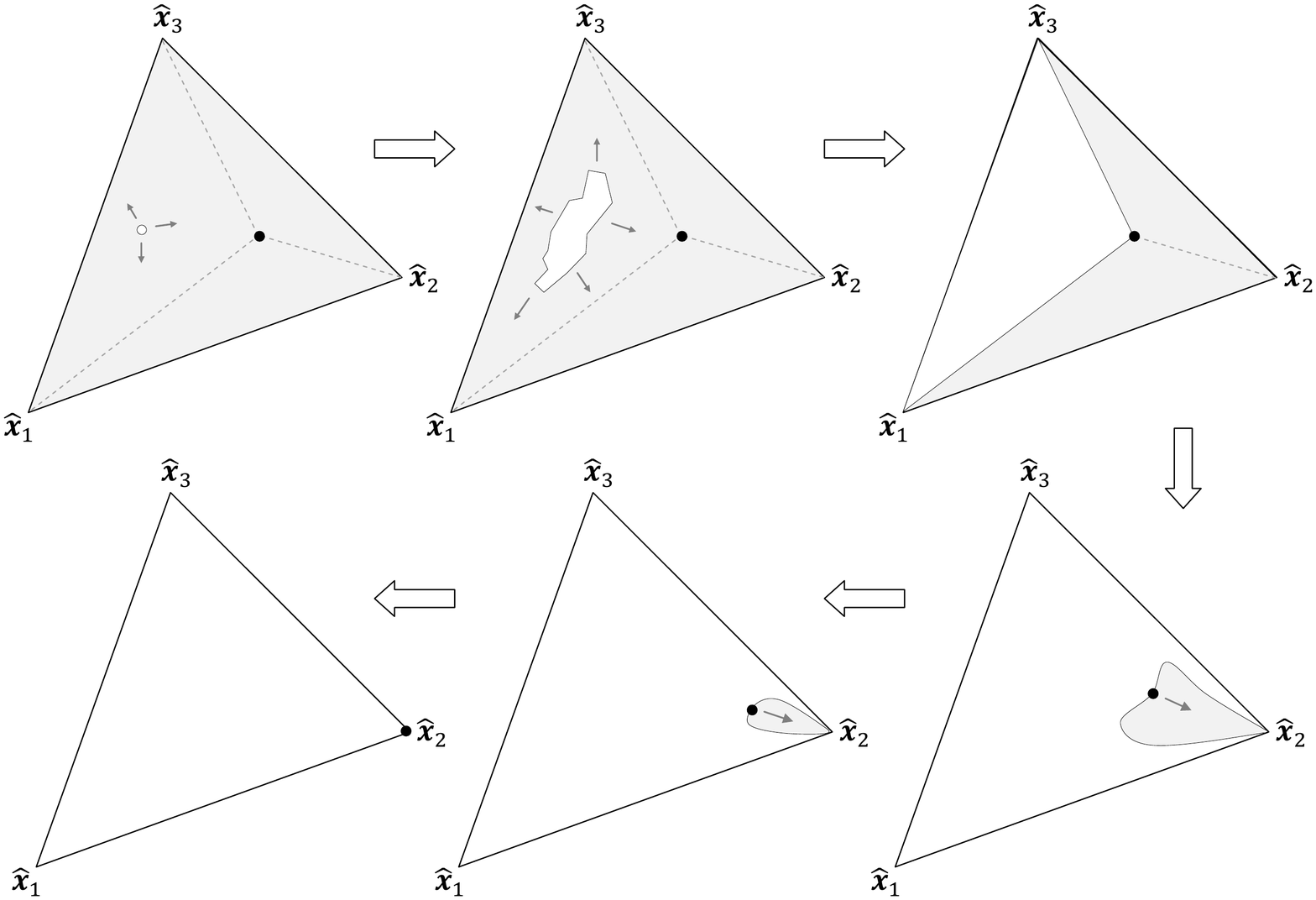}
\caption{The breaking of the $\rho$-elastic membrane (in grey color) in a $e_{\{1\}\{2\}\{3\}}^\rho$ measurement proceeds in two steps: first the membrane collapses, within the boundaries of the convex region containing the initial breaking point (here $A_2$), then, as soon as it looses the anchor points associated with this region, it shrinks towards the remaining anchor point, bringing with it the particle (here to position $\hat{\bf{x}}_2$).  
\label{breakingprocess}}
\end{figure}

Then, after some time the elastic membrane breaks, at some unpredictable point \mbox{\boldmath$\lambda$} (see Fig.~\ref{triangolo}). If $\mbox{\boldmath$\lambda$}\in A_2$,  the tearing  will propagate inside the entire region $A_2$, but not in the other two regions $A_1$ and $A_3$ (due to the presence of the tension lines), causing also its $2$ anchor points $\hat{\bf{x}}_1$ and $\hat{\bf{x}}_3$ to  tear away (from a physical point of view, the ``collapse'' of the membrane in region $A_2$ should be understood as a sort of explosive-like reaction of disintegration of its atomic constituents). Once the membrane is detached from the two above mentioned anchor points, being elastic, it  contracts toward the remaining anchor point $\hat{\bf{x}}_2$, drawing in this way the point particle, which is attached to it, to the same final position (see Fig.~\ref{breakingprocess}).
Similarly, if $\mbox{\boldmath$\lambda$}\in A_1$, the final state of the particle is $\hat{\bf{x}}_1$, and if $\mbox{\boldmath$\lambda$}\in A_3$, the final state of the particle is $\hat{\bf{x}}_3$. 

As for the previous description of the one-dimensional elastic band, we can observe that to each breaking point $\mbox{\boldmath$\lambda$}\in S_2$, it corresponds a specific interaction between the particle and the elastic membrane, almost deterministically drawing the former  to its final state. In other terms, the measurement $e_{\{1\}\{2\}\{3\}}^\rho$ is formed by a collection of \emph{pure measurements}, i.e., potential, almost deterministic measurement interactions, only one of which is each time actualized when the membrane breaks. Again, if we say that the interactions are \emph{almost} deterministic, and not purely deterministic, it is because for those  \mbox{\boldmath$\lambda$}  which are at the boundaries of two (or three) regions, it is not predetermined which region will disintegrate. These special \mbox{\boldmath$\lambda$}, however, are of zero measure in the integrals defining the transition probabilities. 

Following the same logic as for the two-outcome case, the transition probability ${\bf{x}}\to \hat{\bf{x}}_{1}$, for a membrane of the $\rho$-kind, can be written as: 
\begin{eqnarray}
\label{x1integral-triple}
P({\bf{x}}\to \hat{\bf{x}}_{1}|\rho)=\int_{A_1}\rho({\bf{y}})d{\bf{y}} = \int_{A_1}\rho(y_1,y_2, y_3)dy_1dy_2dy_3,
\end{eqnarray}
and similarly for transitions ${\bf{x}}\to \hat{\bf{x}}_{2}$ and ${\bf{x}}\to \hat{\bf{x}}_{3}$. Observing that  $y_1+y_2 + y_3= 1$, we  introduce the new integration variables (associated with a new orthonormal basis $\{\hat{\bf z}_1,\hat{\bf z}_2,\hat{\bf z}_3\}$ in $\mathbb{R}^3$):
\begin{eqnarray}
&z_1={y_3-y_2\over\sqrt{2}}, \quad z_2={2y_1-y_2 -y_3\over\sqrt{6}}, \quad z_3={y_1+y_2 +y_3\over\sqrt{3}}= {1\over\sqrt{3}}\\
&y_1=\sqrt{2\over 3} z_2+ {1\over 3}, \quad y_2 =-{1\over \sqrt{6}}(\sqrt{3} z_1+z_2) + {1\over 3},\quad y_3 ={1\over \sqrt{6}}(\sqrt{3} z_1-z_2)+  {1\over 3},\nonumber
\end{eqnarray}
and transform the triple integral (\ref{x1integral-triple}) into the double integral
\begin{eqnarray}
\label{x1integral-triple-2}
P({\bf{x}}\to \hat{\bf{x}}_{1}|\rho)= \int_{A_1}\rho(z_1,z_2)dz_1dz_2,
\end{eqnarray}
where we have defined: 
\begin{eqnarray}
&\rho(z_1,z_2)\equiv \rho\left(\sqrt{2\over 3} z_2+ {1\over 3}, -{1\over \sqrt{6}}(\sqrt{3} z_1+z_2) + {1\over 3}, {1\over \sqrt{6}}(\sqrt{3} z_1-z_2)+  {1\over 3}\right).
\end{eqnarray}

Considering that the area of $S_2$ is ${\sqrt{3}\over 2}$, for the special case of a uniform probability density (associated with a  membrane made of a uniform material) we have $\rho_u= {2\over\sqrt{3}}$, so that the above integral becomes:
\begin{eqnarray}
\label{x1integral-triple-3}
P({\bf{x}}\to \hat{\bf{x}}_{1}|\rho_u)=  {2\over\sqrt{3}} \int_{A_1}dz_1dz_2.
\end{eqnarray}
Being $A_1$ a triangle of base $\sqrt{2}$ and height $h= \sqrt{3\over 2}x_1$, its area is ${\sqrt{2} h\over 2}={\sqrt{3}\over 2} x_1$. Thus,  in accordance with the analysis of the UTR-model in \citet{AertsSassolideBianchi2014}, we obtain a result in agreement with the Born rule (\ref{Born}), i.e., $P({\bf{x}}\to \hat{\bf{x}}_{1}|\rho_u) =x_1$, and similarly for the probabilities of the other two possible transitions. In other terms, the uniform membrane measurement $e_{\{1\}\{2\}\{3\}}^{\rho_u}$  is  isomorphic to the measurement of an non-degenerate observable $A$, in a three-dimensional complex Hilbert space ${\mathcal H}_3$, if we  represent the quantum state vector $|\psi\rangle = \sqrt{x_1}e^{i\alpha_1}|a_1\rangle +\sqrt{x_2}e^{i\alpha_2}|a_2\rangle +\sqrt{x_3}e^{i\alpha_3}|a_3\rangle \in {\mathcal H}_3$, by a vector ${\bf{x}} = x_1 \hat{\bf{x}}_1 + x_2 \hat{\bf{x}}_2 + x_3 \hat{\bf{x}}_3\in S_2$, whose components are precisely the transition probabilities (see Sec.~\ref{Quantum Probabilities for a Single Observable}).
\\
\vspace{-0.3cm}
\\
\emph{The general $N$-outcome case}
\\
It is straightforward to generalize the working of the GTR-model to the case of an arbitrary number $N$ of outcomes (for the degenerate case, see \citet{AertsSassolideBianchi2014}).  The material point particle then lives in $\mathbb{R}^{n}$, with $n\geq N$,  
and to perform a (non-degenerate) measurement $e_{\{1\}\cdots\{N\}}^{\rho}$, a  $(N-1)$-dimensional hypermembrane of the $\rho$ kind is stretched  over the hypersurface $S_{N-1}$ of a $(N-1)$-dimensional simplex generated by $N$ orthonormal vectors $\hat{\bf x}_1, \dots, \hat{\bf x}_{N}$, and is attached to its $N$ vertex points. Once the  hypermembrane is in place, the particle, by moving deterministically towards it (along a trajectory that is not important here to specify), sticks to it at a particular point: 
\begin{eqnarray}
\label{vector-N}
{\bf x}=\sum_{i\in I_{N}} x_i \hat{\bf x}_i, \quad  \sum_{i\in I_{N}} x_i = 1, \quad I_{N} \equiv \{1,\dots, N\},
\end{eqnarray}
which defines the state of the particle on the hypermembrane.

This gives rise to $N$ ``tension lines,'' connecting  ${\bf x}$ to the different vertex points $\hat{\bf x}_1, \dots, \hat{\bf x}_{N}$, defining in this way $N$ disjoint regions $A_i$, the convex closures of $\{\hat{\bf x}_1, \dots, \hat{\bf x}_{i-1}, {\bf x}, \hat{\bf x}_{i+1}, \dots, \hat{\bf x}_{N}\}$, such that $S_N=\cup_{i\in I_{N}}A_i$. Then, after some time the hypermembrane breaks, at some point ${\mbox{\boldmath$\lambda$}} =\sum_{i\in I_{N}} \lambda_i \hat{\bf x}_i$, $\sum_{i\in I_{N}} \lambda_i = 1$. If $\mbox{\boldmath$\lambda$}\in A_i$, for a given $i\in I_{N}$, then $A_i$ collapses, causing its $N-1$ anchor points $\hat{\bf x}_j$, $j\neq i$, to tear away. So, if  $\mbox{\boldmath$\lambda$}\in A_i$, the elastic $\rho$-hypermembrane  contracts toward point $\hat{\bf x}_{i}$, that is, toward the only point at which it remained attached, pulling in this way  the particle into that position. In other terms, the process produces the transition ${\bf x}\to \hat{\bf x}_{i}$. The probability of such process is: 
\begin{eqnarray}
P({\bf x}\to \hat{\bf x}_{i}|\rho)= \int_{A_i}\rho({\bf y})d{\bf y} = \int_{A_i}\rho(y_1,\dots, y_{N})dy_1\dots dy_{N},
\end{eqnarray}
and exploiting the fact that $\sum_{i=1}^{N} y_i =1$, we can  perform a  suitable change of variables (i.e., of basis in $\mathbb{R}^{N}$) to transform the above $N$-variables integral into a $(N-1)$-variables integral:
\begin{eqnarray}
\label{z-integral-general}
P({\bf x}\to \hat{\bf{x}}_{i}|\rho) = \int_{A_i}\rho({\bf z})d{\bf z}  = \int_{A_i}\rho(z_1,\dots, z_{N-1})dz_1\dots dz_{N-1},
\end{eqnarray}
where the constant variable is $z_{N}={1\over\sqrt{N}}\sum_{i=1}^{N} y_i ={1\over\sqrt{N}}$. It is not difficult to show that in the uniform case one obtains, for all $i\in I_N$ (see the proof in \citet{AertsSassolideBianchi2014}),
\begin{eqnarray}
\label{transitionprobabilitiesnondeg}
P({\bf x}\to \hat{\bf{x}}_{i}|\rho_u) = x_i, 
\end{eqnarray}
showing that the measurement  $e_{\{1\}\cdots\{N\}}^{\rho_u}$  is  isomorphic to the measurement of a non-degenerate observable (\ref{observable-hilbert}), in a $N$-dimensional complex Hilbert space ${\mathcal H}_N$, if we  represent the quantum state vector (\ref{state-hilbert}) by the vector (\ref{vector-N}), whose components are precisely the transition probabilities (\ref{transitionprobabilitiesnondeg}).

\vspace{-0.4cm}
\section{Universal measurements}
\label{theorem}
\vspace{-0.4cm}

The GTR-model immediately suggests the possibility of considering a much more general typology of measurement, expressing a deeper level of potentiality. Indeed, a measurement $e_{\{1\}\cdots\{N\}}^{\rho}$ is a conditional measurement, i.e., a measurements whose outcomes are conditional to the specific choice made for the probability density $\rho$. However, it is natural to assume that in many measurement situations the probability density $\rho$ is not  given a priori, but randomly selected within the infinite  ensemble of all possible $\rho$. The measurements  would then be the result of a  two-level process of actualization of potentials, the first level corresponding to the random choice of a given probability density $\rho$, and the second level to the selection, from that actualized $\rho$, of a deterministic interaction, which finally produces the outcome. 

We thus need to modelize a measurement situation which consists in a meta-measurement, such that the statistics of outcomes are also the result of a process of randomization over different probability densities $\rho$. We call these more general measurements \emph{universal measurements}. Since they involve an average over the non-denumerable set of $N$-dimensional integrable generalized functions $\rho$, one needs to take care, in their definition, not to be confronted with technical problems related to the foundations of mathematics and probability theory (see the discussion in \citet{AertsSassolideBianchi2014}). This can be done by adopting the following strategy: 

\vspace{0.1cm}
\noindent (1) First, one shows that any  probability density $\rho$ can be described as the limit of a suitably chosen sequence of \emph{cellular probability densities} $\rho_{n_c}$, as the number of cells $n_c$ tends to infinity, in the sense that for every initial state ${\bf x}$ and final state $\hat{\bf{x}}_{i}$, $i\in I_N$, we can always find a sequence of cellular $\rho_{n_c}$, such that the transition probability $P({\bf x}\to\hat{\bf{x}}_{i}|\rho_{n_c})$ tends to $P({\bf x}\to \hat{\bf{x}}_{i}|\rho)$, as $n_c\to\infty$. By a cellular probability density we mean a probability density describing a structure made of a total number $n_c$ of regular cells (of whatever shape), which  tessellate the hypersurface of the simplex $S_{N-1}$. These $n_c$ cells can only be of two sorts: such that $\rho_{n_c}$ is equal to a constant inside them (the same constant for all cells), or such that $\rho_{n_c}$ is equal to zero inside them, which in the physical realization in terms of hypermembranes corresponds, respectively, to the situation of uniformly breakable cells and unbreakable cells.

\vspace{0.1cm}
\noindent (2) Thanks to the fact that a cellular probability density $\rho_{n_c}$ is only made of a finite number $n_c$ of cells, which can either be of the breakable or unbreakable kind, if we exclude the totally unbreakable case of a $\rho_{n_c}$ describing a structure only made of unbreakable cells (the trivial case $\rho_{n_c}\equiv 0$, producing no outcomes in a measurement), we have that, given a  $n_c \in \mathbb{N}$, the total number of possible $\rho_{n_c}$ is $C_{n_c}^0 + C_{n_c}^1 + C_{n_c}^2 + \cdots + C_{n_c}^{n_c} -1 = 2^{n_c}-1$. Therefore, for each $n_c$, we can unambiguously define the average probability: 
\begin{eqnarray}
\label{average1}
P({\bf x}\to \hat{\bf{x}}_{i}|n_c) \equiv {1\over 2^{n_c}-1}  \sum_{\rho_{n_c}}P({\bf x}\to \hat{\bf{x}}_{i}|\rho_{n_c}),
\end{eqnarray}
where the sum runs over all the possible $2^{n_c}-1$ (non-zero) cellular probability densities $\rho_{n_c}$, made of $n_c$ cells. 

Clearly, $P({\bf x}\to \hat{\bf{x}}_{i}|n_c)$ is the probability of transition ${\bf x}\to \hat{\bf{x}}_{i}$, when a cellular hypermembrane $\rho_{n_c}$ is chosen at random, in a uniform way. The uniform average (\ref{average1}), being over a finite number of $\rho_{n_c}$, is uniquely defined and doesn't suffer from possible ``Bertrand paradox'' ambiguities. Also, considering point (1) above, i.e., the fact that the $\rho_{n_c}$ are dense in the space of probability densities (in the sense specified above), we are in a position to give the following general definition of a universal measurement.\\

\vspace{-0.4cm}
\noindent {\bf Definition (Universal Measurement)}. \emph{A measurement $e^{\rm{univ}}_{\{1\}\cdots\{N\}}$ is said to be a \emph{universal measurement} if the probabilities associated with all its $N$ possible transitions ${\bf x}\to \hat{\bf{x}}_{i}$, $i\in I_N$, are the result of a uniform average over all possible measurements $e^\rho_{\{1\}\cdots\{N\}}$, described by all possible probability densities $\rho$, as defined by the infinite-cell limit: 
\begin{eqnarray}
\label{average-limit}
P^{\rm{univ}}({\bf x}\to \hat{\bf{x}}_{i})=\lim_{n_c\to\infty} P({\bf x}\to \hat{\bf{x}}_{i}|n_c),
\end{eqnarray}
where $P({\bf x}\to \hat{\bf{x}}_{i})|n_c)$ is the average  \emph{(\ref{average1})}. 
}
\\
\vspace{-0.4cm}

We can observe that in order to define a uniform randomization over all possible $\rho$, i.e., a probability measure over the integrable (generalized) functions $\rho$, without being confronted with insurmountable technical problems  related to the foundations of mathematics, we have followed here a strategy which is similar to what is done in the definition of the \emph{Wiener measure}, which is a  probability law on the space of continuous functions, describing for instance the Brownian motions. As is well known, the Wiener measure allows to attribute probabilities to continuous-time random walks, and it succeeds to do so by considering them as the limit of discrete-time processes. In the analysis of Brownian processes, one starts with the description of particles that can move only on regular cellular structures (regular lattices), which at each step can jump from one location to another, according to a given probability law, and then consider the (continuous) limit where these steps become infinitesimal. In our definition of universal measurements, we have proceeded in a similar logic, by first considering discretized structures $\rho_{n_c}$, and their uniform average, which is perfectly well defined, taking then in the end the infinite limit $n_c\to\infty$.

The following theorem establishes the connection between universal measurements and uniform measurements:\\

\vspace{-0.4cm}
\noindent {\bf Theorem (Universal $\Leftrightarrow$ Uniform)}. \emph{A universal measurement $e^{\rm{univ}}_{\{1\}\cdots\{N\}}$ is probabilistically equivalent to a  measurement $e^{\rho_u}_{\{1\}\cdots\{N\}}$, defined in terms of a uniform probability density $\rho_u$, in the sense that for all transitions ${\bf x}\to \hat{\bf{x}}_{i}$, $i\in I_N$, we have the equality:
\begin{eqnarray}
\label{theoremuniversal}
P^{\rm{univ}}({\bf x}\to \hat{\bf{x}}_{i})=P({\bf x}\to \hat{\bf{x}}_{i}|\rho_u).
\end{eqnarray}
}
\\
\vspace{-0.4cm}

To prove the theorem, we proceed with the following steps. First, in Sec.~\ref{limit of cellular structures}, we show that one can always consider limits of cellular $\rho_{n_c}$, to approximate the transition probabilities associated with arbitrary $\rho$ (also including the possibility of Dirac distributions), so that the average (\ref{average1}) does actually include all possible $\rho$, as $n_c\to\infty$, i.e., it is a universal average. Then, in Sec.~\ref{Averaging over finite cellular structures}, we study the average probability over all possible kinds of cellular structures $\rho_{n_c}$, for a fixed $n_c$, and show, by a recurrence method, that for all $n_c\in \mathbb{N}$ and $i\in I_N$:
\begin{eqnarray}
\label{averagerho2}
P({\bf x}\to \hat{\bf{x}}_{i}|n_c) = {1\over 2^{n_c}-1}  \sum_{\rho_{n_c}}P({\bf x}\to \hat{\bf{x}}_{i}|\rho_{n_c})=P({\bf x}\to \hat{\bf{x}}_{i}|\rho_{u;n_c}),
\end{eqnarray}
where $\rho_{u;n_c}$ is the  probability density describing a uniformly breakable structure made of $n_c$-cells. Then, considering that $P({\bf x}\to \hat{\bf{x}}_{i}|\rho_{u;n_c})\to P({\bf x}\to \hat{\bf{x}}_{i}|\rho_{u})$, as $n_c\to\infty$, (\ref{averagerho2})  proves (\ref{theoremuniversal}).

\vspace{-0.4cm}
\section{Limits of cellular structures}
\label{limit of cellular structures}
\vspace{-0.4cm}

In this section we show that an arbitrary measurement $e^\rho_{\{1\}\cdots\{N\}}$ can always be probabilistically described in terms of measurements $e^{\rho_{n_c}}_{\{1\}\cdots\{N\}}$, performed by means of cellular hypermembranes $\rho_{n_c}$, made of $n_c$ elementary cells, in the limit $n_c\to\infty$. By this we mean that, given a probability density $\rho$ on a simplex $S_{N-1}$, we can always find a suitable sequence of cellular hypermembranes  $\rho_{n_c}$, such that $P({\bf x}\to\hat{\bf{x}}_{i}|\rho_{n_c}) \to P({\bf x}\to \hat{\bf{x}}_{i}|\rho)$, as $n_c\to\infty$, for all $i\in I_N$.
\\
\vspace{-0.3cm}
\\
\emph{The $N=2$ case (two outcomes)}
\\
We start by proving the result in the case of a one-dimensional probability density (describing, in the physical realization of the model, a one-dimensional elastic band); we will then show that the proof straightforwardly generalizes to higher dimensional systems. 

Our goal is to show that to each one-dimensional probability density $\rho(z)$, on the line segment $[-{1\over\sqrt{2}}, {1\over\sqrt{2}}]$ (corresponding to the 1-simplex $S_1$), it is always possible to find a suitable sequence of cellular distributions $\rho_{n_c}(z)$, such that:  
\begin{eqnarray}
\label{limitntoinfty}
\lim_{n_c\to\infty}\left[ P({\bf x}\to \hat{\bf x}_{i}|\rho) - P({\bf x}\to \hat{\bf x}_{i}|\rho_{n_c}) \right]=0,
\end{eqnarray}
with $i=1,2$. For this, we partition the interval $[-{1\over\sqrt{2}}, {1\over\sqrt{2}}]$ into $n_c=m\ell$ elementary intervals: $[-{1\over\sqrt{2}},-{1\over\sqrt{2}}+{\sqrt{2}\over n_c}]$, $[-{1\over\sqrt{2}}+{\sqrt{2}\over n_c},-{1\over\sqrt{2}}+2{\sqrt{2}\over n_c}],\dots,[-{1\over\sqrt{2}}+(i-1){\sqrt{2}\over n_c},-{1\over\sqrt{2}}+i{\sqrt{2}\over n_c}],\dots,[{1\over\sqrt{2}}-{\sqrt{2}\over n_c},{1\over\sqrt{2}}]$. These elementary intervals (i.e., elementary one-dimensional cells), of length ${\sqrt{2}\over n_c}$, are in turn contained in $m={n_c\over \ell}$ larger intervals, of length ${\sqrt{2}\over m} = {\sqrt{2}\ell\over n_c}$, which are the following: $[-{1\over\sqrt{2}},-{1\over\sqrt{2}}+ {\sqrt{2}\over m}], [-{1\over\sqrt{2}}+ {\sqrt{2}\over m},-{1\over\sqrt{2}}+2 {\sqrt{2}\over m}],\dots, [-{1\over\sqrt{2}}+(i-1) {\sqrt{2}\over m},-{1\over\sqrt{2}}+i {\sqrt{2}\over m}],\dots, [{1\over\sqrt{2}}- {\sqrt{2}\over m},{1\over\sqrt{2}}]$. In other terms, denoting
\begin{eqnarray}
&S_i\equiv \left[-{1\over\sqrt{2}}+(i-1) {\sqrt{2}\over m},-{1\over\sqrt{2}}+i {\sqrt{2}\over m}\right],\\
&\sigma_{i,j}\equiv \left[-{1\over\sqrt{2}}+(i-1) {\sqrt{2}\over m}+(j-1){\sqrt{2}\over n_c},-{1\over\sqrt{2}}+(i -1){\sqrt{2}\over m}+j{\sqrt{2}\over n_c}\right], 
\end{eqnarray}
we have: $S_i=\bigcup_{j=1}^{\ell} \sigma_{i,j}$, $\left[-{1\over\sqrt{2}}, {1\over\sqrt{2}}\right] =\bigcup_{i=1}^{m} S_i = \bigcup_{i=1}^{m}\bigcup_{j=1}^{\ell} \sigma_{i,j}$.

We assume that $x_1$ is such that $-{1\over\sqrt{2}}(1-2x_1)\in (-{1\over\sqrt{2}}+(j-1) {\sqrt{2}\over m},-{1\over\sqrt{2}}+j {\sqrt{2}\over m}]$, for some given $j$. This condition can also be expressed as $mx_1\in (j-1,j]$. Therefore, by definition of the ceiling function, $\lceil mx_1\rceil=j$, and we can write: 
%Massimiliano: in the formulae below I have corrected a few minor things, as I did in our proof in the other papers
\begin{eqnarray}
\label{partition4}
P({\bf x}\to \hat{\bf x}_{1}|\rho)= \int_{-{1\over\sqrt{2}}}^{-{1\over\sqrt{2}}(1-2x_1)} \rho(z) dz=\sum_{i=1}^{\lceil mx_1\rceil-1}\int_{S_i} \rho(z)dz +r_{m}(x_1|\rho),
\end{eqnarray}
where the rest:
\begin{eqnarray}
\label{restrho}
r_{m}(x_1|\rho)= \int_{-{1\over\sqrt{2}}[1-(\lceil mx_1\rceil -1){2\over m}]}^{-{1\over\sqrt{2}}(1-2x_1)}\rho(z)dz
\end{eqnarray}
tends to zero as $m\to\infty$, considering that 
\begin{eqnarray}
\label{ceilinglimit}
\lim_{m\to\infty}{\lceil mx_1\rceil \over m}=x_1.
\end{eqnarray}

At this point, we introduce the following cellular probability density ($n_c =m\ell$):
\begin{eqnarray}
\label{cellular}
\rho_{m\ell}(z)={\chi_{m\ell}(z)\over \int_{-{1\over\sqrt{2}}}^{1\over\sqrt{2}}\chi_{m\ell}(z)dz} ={\chi_{m\ell}(z)\over n_c^b {\sqrt{2}\over n_c}},
\end{eqnarray}
describing a cellular elastic band made of $n_c=m\ell$ elementary cells, which can only be of two sorts: breakable or unbreakable. Here $\chi_{m\ell}(z)$ denotes a  step-like function, taking the constant values $1$ (for the breakable cells) or $0$ (for the unbreakable cells) inside each interval $\sigma_{i,j}$, and $n^b_c$ is the total number of breakable elementary cells of the structure. For a cellular probability density of this kind, (\ref{partition4})  becomes:
\begin{eqnarray}
\label{partition5}
P({\bf x}\to \hat{\bf x}_{1}|\rho_{m\ell})= \int_{-{1\over\sqrt{2}}}^{-{1\over\sqrt{2}}(1-2x_1)} \rho_{m\ell}(z) dz=\sum_{i=1}^{\lceil mx_1\rceil -1}{n_c^{b,i}\over n_c^b} +r_{m}(x_1|\rho_{m\ell}),
\end{eqnarray}
where $n_c^{b,i}$ is the number of breakable cells in $S_i$, and the rest is defined as in (\ref{restrho}). Comparing (\ref{partition4}) with (\ref{partition5}), we obtain:
\begin{eqnarray}
\lefteqn{P({\bf x}\to \hat{\bf x}_{1}|\rho)-P({\bf x}\to \hat{\bf x}_{1}|\rho_{m\ell})=\int_{-{1\over\sqrt{2}}}^{-{1\over\sqrt{2}}(1-2x_1)} [\rho(z)- \rho_{m\ell}(z)]dz}\nonumber\\
\label{difference-m-l}
&&= \sum_{i=1}^{\lceil mx_1\rceil -1}\left(\int_{S_i} \rho(z)dz - {n_c^{b,i}\over n_c^b}\right)+[r_{m}(x_1|\rho)-r_{m}(x_1|\rho_{m\ell})].
\end{eqnarray}

All we need to do is to observe that we can always choose $\rho_{m\ell}(z)$ in such a way that ${n_c^{b,i}\over n_c^b}\to\int_{S_i} \rho(z)dz$, as $\ell\to\infty$, for all $i=1,\dots,m$. This because, for each $i$, the probability $\int_{S_i} \rho(z)dz$ is a real number in the interval $[0,1]$, and rational numbers of the form ${n_c^{b,i}\over n_c^b}$, with $0\leq n_c^{b,i}\leq n_c^b$, $n_c^b>0$, are dense in $[0,1]$. Therefore, for such a choice of $\rho_{m\ell}(z)$, taking  the limit $\ell\to\infty$, the sum in (\ref{difference-m-l}) vanishes, and taking the limit $m\to\infty$, also the two rests in (\ref{difference-m-l}) vanish, so that we can conclude that (\ref{limitntoinfty}) holds, i.e., that we can always find a suitable sequence of probability densities $\rho_{n_c}\equiv \rho_{m\ell}$, describing  structures made of breakable and unbreakable elementary cells, such that in the infinite-cell limit $n_c\to\infty$ they produce exactly the same probabilities as $\rho$. Of course, the same reasoning holds true for outcome $x_2$, considering also that $P({\bf x}\to \hat{\bf x}_{2}|\rho)= 1-P({\bf x}\to \hat{\bf x}_{1}|\rho)$.
\\
\vspace{-0.3cm}
\\
\emph{The general case}
\\
It is straightforward to generalize the above result to the case of a $(N-1)$-dimensional hypermembrane describing a system with $N$ different possible outcomes. For this, we observe that it is always possible to replace in the integral (\ref{z-integral-general}) the probability density $\rho$, defined on $S_{N-1}$, by an extended probability density: 
\begin{equation}
{\tilde \rho}({\bf z}) = \left\{ \begin{array}{ll}
         \rho({\bf z}) & \mbox{if ${\bf z}\in S_{N-1}$}\\
        0 & \mbox{if ${\bf z}\in R_{N-1}\setminus S_{N-1}$},\end{array} \right.
\end{equation}
defined on a $(N-1)$-dimensional hyperrectangle $R_{N-1}$, which we assume is oriented according to the ${\bf z}$-coordinate system, and is large enough to contain the simplex $S_{N-1}$. In other terms, we can  write:
\begin{eqnarray}
\label{integral-N-variables-bis}
P({\bf x}\to \hat{\bf x}_{i}|\rho) =P({\bf x}\to \hat{\bf x}_{i}|\tilde \rho)=\int_{A_i}{\tilde \rho}(z_1,\dots, z_{N-1})dz_1\dots dz_{N-1},
\end{eqnarray}
with the integrand now defined on the entire hyperrectangle $R_{N-1}$. 

Writing $R_{N-1}$ as the Cartesian product of one dimensional intervals, i.e., $R_{N-1} = I_1\times I_2\cdots I_{N-2}\times I_{N-1}$, where $I_k = [a_k,b_k]$, $k=1,\dots,N-1$, we can partition each one of these intervals as follows:
\begin{eqnarray}
&I_k =\bigcup_{i=1}^{m} S_{i;k} = \bigcup_{i=1}^{m}\bigcup_{j=1}^{\ell} \sigma_{i,j;k},\quad  S_{i;k}\equiv \left[a_k+(i-1) {b_k-a_k\over m}, a_k+i {b_k-a_k\over m}\right],\nonumber\\
&\sigma_{i,j;k}\equiv
\left[a_k+(i-1) {b_k-a_k\over m}+(j-1) {b_k-a_k\over n},a_k+(i-1) {b_k-a_k\over m}+j {b_k-a_k\over n}\right],\nonumber
\end{eqnarray}
with $n=m\ell$. In this way, we  obtain the tessellation
\begin{eqnarray}
\lefteqn{R_{N-1}= \bigcup_{i_1,\dots, i_{N-1}=1}^{m} S_{i_1;1}\times\cdots\times S_{i_{N-1};N-1}}\nonumber\\
&&=\bigcup_{i_1,\dots, i_{N-1}=1}^{m} \bigcup_{j_1,\dots, j_{N-1}=1}^{\ell} \sigma_{i_1,j_1;1}\times\cdots\times \sigma_{i_N,j_N;N}.
\end{eqnarray} 

Similarly to what we have done for the one-dimensional case, we can then introduce a cellular probability density $\tilde\rho_{n_c}$, taking constant values $1$ or $0$ on each of the $n_c= n^{N-1} = (m\ell)^{N-1}$ elementary hyperrectangles $\sigma_{i_1,j_1;1}\times\cdots\times \sigma_{i_{N-1},j_{N-1};N-1}$, and study  the probability difference $P({\bf x}\to \hat{\bf x}_{i}|\tilde\rho)-P({\bf x}\to \hat{\bf x}_{i}|\tilde\rho_{n_c})$, $i\in I_N$. Here we have to observe that the convex regions of integration $A_i$ only intersect a  finte number of $(N-1)$-dimensional hyperrectangular cells $S_{i_1;1}\times\cdots\times S_{i_{N-1};N-1}$, so that the indices associated with these ``peripheral'' cells define, for each integration variable, the higher and lower values for the summation indices, when the multiple integral of the probability difference is written as a sum of integrals over these hyperrectangular cells. This  produces a finite number of rests, which  tends to zero when $m$ goes to infinity. We also observe, similarly to the one-dimensional case, that the integrals of  $\tilde\rho_{n_c}$ over the cells $S_{i_1;1}\times\cdots\times S_{i_{N-1};N-1}$ are equal to the number of breakable elementary cells they contain, divided by the  number of breakable elementary cells of the entire structure. Therefore, by letting $\ell\to\infty$, and by suitably choosing $\tilde\rho_{n_c}$, we can  always make sure that these rational numbers tend to the real numbers corresponding the integrals associated with $\tilde\rho$. Observing that $\rho\equiv\tilde\rho$ on $S_{N-1}$, we  thus conclude that (\ref{limitntoinfty}) holds also in the $(N-1)$-dimensional case, i.e., when $\rho$ is a $(N-1)$-dimensional hypermembrane describing a measurement with $N$ different outcomes.

\vspace{-0.4cm}
\section{Averaging over cellular structures}
\label{Averaging over finite cellular structures}
\vspace{-0.4cm}

In the previous section we have shown that a measurement with an arbitrary $\rho$ (an arbitrary hypermembrane) can always be understood as the  limit of measurements performed by discretized structures $\rho_{n_c}$, having a finite number ${n_c}$ of breakable and unbreakable cells, when the number of these cells tends to infinity. We want now to determine the probabilities of measurements describing situations where the cellular structure $\rho_{n_c}$ is not given a priori, but selected at random among all possible structures with a given  ${n_c}$. We  start considering the one-dimensional case (two outcomes), then will show that the analysis straightforwardly generalize to an arbitrary number of dimensions (i.e., of outcomes).
\\
\vspace{-0.3cm}
\\
\emph{The one-dimensional case}
\\
We consider one-dimensional cellular elastic structures. As already mentioned, cells can only be of two sorts: breakable ($b$) or unbreakable ($u$). For simplicity, we also assume that the point particle can only be located on the elastic in a position corresponding to the contact point between two cells (in other terms, we exclude the two end points of the elastic, as in any case they correspond to eigenstates, which are not affected by the measurements). 

\begin{figure}[!ht]
\centering
\includegraphics[scale =.9]{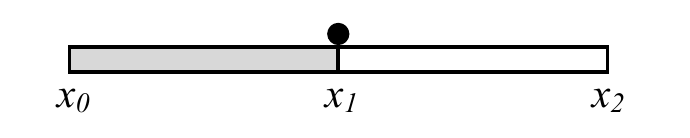}
\caption{A $2$-cell structure in the ``left breakable'' configuration $(bu)$. The breakable cell is represented in grey color, the unbreakable one in white color. The material point particle occupies the only available position between the two cells.
\label{2-cells}}
\end{figure}

The simplest non trivial case is when the elastic is made of two cells. Then, the number of  different possible elastic structures, excluding the totally unbreakable one, is  $2^2 -1  = 3$, which are: $(bb)$ = uniformly breakable, $(ub)$ = right breakable, and $(bu)$ = left breakable. For a two-cell elastic, in addition to its two end points, which we  denote $x_0$ (left end) and $x_2$ (right end), there is a unique internal point of contact between the two cells (see Fig.~\ref{2-cells}), which we denote $x_1$. This is the only point that can be occupied by the particle, and which can give rise to the transition $x_1\to x_0$, if the right cell breaks, or $x_1\to x_2$, if the left cell breaks.

We simply denote $P(i\to j|xy)$ the probability of  transition $x_i\to x_j$, knowing that the elastic is of the $(xy)$ kind, $x,y\in\{b,u\}$. For transition $1\to 0$ (i.e., $x_1\to x_0$), we  have: 
\begin{eqnarray}
&P(1\to 0|bb) =   {1\over 2}, \quad P(1\to 0|ub) = 1, \quad P(1\to 0|bu) = 0,
\end{eqnarray}
whereas for transition $1\to 2$ (i.e., $x_1\to x_2$) we have:
\begin{eqnarray}
&P(1\to 2|bb) = {1\over 2}, \quad  P(1\to 2|ub) = 0, \quad P(1\to 2|bu) = 1.
\end{eqnarray}

We now  assume that the elastic is not a priori given, but chosen each time at random, in a uniform way, i.e., in a way such that all the elastics have exactly the same chance to be selected. Then, if $\mu(xy)$ denotes the probability that elastic $(xy)$ is chosen, we have $\mu(xy)={1\over 3}$. If we  denote $P(1\to j|2)$ the probability of  transition $x_1\to x_j$, when the 2-cell elastic is chosen in a uniformly random way (excluding from the choice the $(uu)$ fully unbreakable structure), by definition we have: 
\begin{eqnarray}
P(1\to j|2) = \sum_{(xy)} \mu(xy) P(1\to j|xy) = {1\over 3} \sum_{(xy)} P(i\to j|xy).
\end{eqnarray}
For  transition $1\to 0$, we therefore obtain:
\begin{eqnarray}
P(1\to 0|2) &=& {1\over 3}\left[P(1\to 0|bb)+P(1\to 0|ub)+  P(1\to 0|bu) \right]\nonumber\\
&=& {1\over 3}\left({1\over 2}+ 1+0\right) = {1\over 3}{3\over 2}={1\over 2},
\end{eqnarray}
and of course we also have $P(1\to 2|2)= 1- P(0\to 2|2)= {1\over 2}$. Thus, we find that the transition probabilities for a randomly chosen breakable elastic, are identical to the transition probabilities associated with the uniformly breakable elastic: 
\begin{eqnarray}
P(1 \to j|2)=P(1\to j|bb), \quad j\in\{0,2\}.
\end{eqnarray}

Having explicitly checked the $2$-cell 
%Massimiliano: and $3$-cell examples
example, we want now to prove that:
\begin{eqnarray}
P(i \to j|n)=P(i\to j|\underbrace{b\cdots b}_n), \quad i\in\{1,\dots,n-1\}, j\in\{0,n\},
\end{eqnarray}
remains true also in the general $n$-cell case. The total number of different breakable elastics is now $2^n-1$, and the uniform measure over elastics is   $\mu(x\cdots)={1\over 2^n -1}$. Thus:
\begin{eqnarray}
P(i \to j|n) = {1\over 2^n-1}\sum_{(x\cdots)} P(i\to j|x\cdots),
\end{eqnarray}
and what we need  to prove is that:
\begin{eqnarray}
\sum_{(x\cdots)} P(i\to j|x\cdots) = (2^n-1)P(i\to j|b\cdots b).
\end{eqnarray}
With no loss of generality, we limit our discussion to the case $j=0$. Then, since $P(i\to 0|b\cdots b)={n-i\over n}$, the above equality becomes:
\begin{eqnarray}
\label{equality-i}
\sum_{(x\cdots)} P(i\to 0|x\cdots) = (2^n-1){n-i\over n},
\end{eqnarray}
and for $i=1$, we have:
\begin{eqnarray}
\label{i=1}
\sum_{(x\cdots)} P(1\to 0|x\cdots) = (2^n-1){n-1\over n}.
\end{eqnarray}

We start by writing:
\begin{eqnarray}
\sum_{(x\cdots)} P(1\to 0|x\cdots) =\sum_{(u\cdots)} P(1\to 0|u\cdots)+ \sum_{(b\cdots )} P(1\to 0|b\cdots ),
\end{eqnarray}
where the first sum in the r.h.s. of the equation runs over all $n$-cell elastics starting with a left unbreakable cell, and the second sum runs over all $n$-cell elastics starting with a left breakable cell. We can observe that all probabilities in the first sum are equal to $1$, so that the sum is equal to $2^{n-1}-1$. Also, the second sum can be written as $\sum_{k=0}^{n-1} {k\over k+1} {n-1 \choose k}$. Using a symbolic computational program (like Mathematica, of Wolfram Research, Inc.), one easily obtains the exact identity:
\begin{eqnarray}
\label{Wolfram}
\sum_{k=0}^{n} {k\over k+1} {n \choose k} = {2^n (n-1) +1\over n+1},
\end{eqnarray}
so that:
\begin{eqnarray}
\sum_{(x\cdots)} P(1\to 0|x\cdots) =2^{n-1}-1 +  {2^{n-1} (n-2) +1\over n}= (2^n-1){n-1\over n},
\end{eqnarray}
which proves (\ref{i=1}). 

To prove (\ref{equality-i}) for an arbitrary $i\in\{1,\dots,n-1\}$, we can reason by recurrence. We have shown that the equality holds for $i=1$; let us assume it holds for some $i$, and that this implies it also holds for $i+1$. We write:
\begin{eqnarray}
\label{sumgeneral}
\lefteqn{\sum_{(x \cdots)} P(i+1\to 0|x\cdots)}\nonumber\\
&&=\sum_{(\cdots u\cdots)} P(i+1\to 0|\cdots u\cdots)+ \sum_{(\cdots b\cdots )} P(i+1\to 0|\cdots b\cdots ),
\end{eqnarray}
where the first sum, in the r.h.s. of the equation, runs over all elastics having an unbreakable $(i+1)$-th cell, and the second sum  runs over all $n$-cell elastics having a breakable $(i+1)$-th cell. Observing that $P(i+1\to 0|\cdots u\cdots)=P(i\to 0|\cdots u\cdots)$, we can write for the first sum:
\begin{eqnarray}
\label{sum-final}
\lefteqn{\sum_{(\cdots u\cdots)} P(i+1\to 0|\cdots u\cdots)=\sum_{(\cdots u\cdots)} P(i\to 0|\cdots u\cdots)}\nonumber\\
&\!\!\!\!=&\!\!\sum_{(\cdots u\cdots)} P(i\to 0|\cdots u\cdots)+\!\!\sum_{(\cdots b\cdots)} P(i\to 0|\cdots b\cdots)-\!\!\sum_{(\cdots b\cdots)} P(i\to 0|\cdots b\cdots)\nonumber\\
&\!\!\!\!=&\sum_{(x\cdots)} P(i\to 0|x\cdots)-\sum_{(\cdots b\cdots)} P(i\to 0|\cdots b\cdots)\nonumber\\ 
&\!\!\!\!=&  (2^n-1){n-i\over n}-\sum_{(\cdots b\cdots)} P(i\to 0|\cdots b\cdots),
\end{eqnarray}
where for the equality of the second line we have added and subtracted the same quantity, and for the last equality we have used (\ref{equality-i}) and the recurrence hypothesis. Then, (\ref{sumgeneral}) becomes:
\begin{eqnarray}
\label{sum-general2}
&&\sum_{(x \cdots)} P(i+1\to 0|x\cdots) =(2^n-1){n-i\over n} +\nonumber\\
&+&\sum_{(\cdots b\cdots)} \left[ P(i+1\to 0|\cdots b\cdots) - P(i\to 0|\cdots b\cdots)\right].
\label{difference}
\end{eqnarray}

Denoting $k_i$ the number of breaking cells at the right of the $i$-th cell, and $k$ the total number of breaking cells, for an elastic of the $(\cdots b\cdots)$ kind, we have $P(i\to 0|\cdots b\cdots)={k_i\over k}$, and $P(i+1\to 0|\cdots b\cdots)={k_i-1\over k}$, so that the difference of probabilities in (\ref{difference}) is equal to $-{1\over k}$, and is independent of $k_i$. Using the exact identity (which again, can be easily obtained using a symbolic computational program, like Mathematica, of Wolfram Research, Inc.):
\begin{eqnarray}
\label{Wolfram3}
\sum_{k=0}^{n} {1\over k+1} {n \choose k}={2^{n+1}-1\over n+1},
\end{eqnarray}
we obtain
\begin{eqnarray}
\label{sumgeneral3}
-\sum_{(\cdots b\cdots)} {1\over k(\cdots b\cdots)} =-\sum_{k=0}^{n-1} {1\over k+1} {n-1 \choose k}=-{2^{n}-1\over n},
\end{eqnarray}
and inserting (\ref{sumgeneral3}) into (\ref{sum-general2}), we finally obtain:
\begin{eqnarray}
\label{final}
\sum_{(x \cdots)} P(i+1\to 0|x\cdots) =(2^n-1){n-i\over n}-{2^{n}-1\over n}=(2^n-1){n-(i+1)\over n},\nonumber
\end{eqnarray}
which proves that (\ref{equality-i}) also holds for $i+1$, thus completing the recurrence proof. 
\\
\vspace{-0.3cm}
\\
\emph{The multidimensional case}
\\
The above demonstration was only for one-dimensional cellular structures, but it is straightforward to generalize it to the case of an arbitrary number of dimensions. Also in this case, the breakable $(b)$ and unbreakable $(u)$ regular cells tessellating the hypermembranes, although  now multidimensional (for instance, in two dimensions, they can be triangles, rectangles, or hexagons), they remain finite in number, so that for a given $n_c$, there is always only a total number  $2^{n_c} -1$ of different possible cellular $n_c$-cell hypermembranes $\rho_{n_c}$.

If we consider an arbitrary region $A$ (not necessarily convex) of a given hypermembrane, then the probability $P(A|\rho_{n_c})$ that one of the cells in $A$ breaks is  given by the ratio between the number of breakable cells in $A$ and the total number of breakable cells in the hypermembrane. Of course, in case $\rho_{n_c}\equiv \rho_{u;n_c}$, i.e., in case the hypermembrane is a uniform structure, made only of breakable cells,  $P(A|\rho_{u;n_c})$ is simply the ratio between the number of cells in $A$ and the number $n_c$ of cells forming the entire hypermembrane. So, if we denote by $i$ the number of cells contained in the complementary region of $A$,  then $n_c-i$ is the number of cells contained in $A$, and we can write:
\begin{eqnarray}
P(A|\rho_{u;n_c})={n_c-i\over n_c}.
\end{eqnarray}

\begin{figure}[!ht]
\centering
\includegraphics[scale =1]{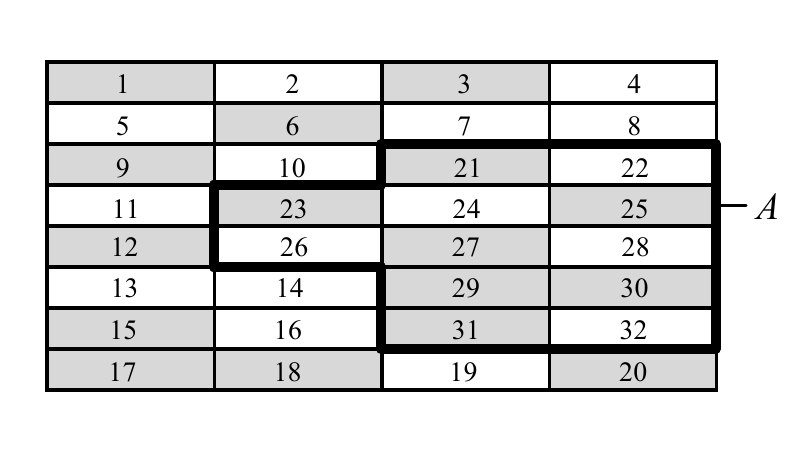}
\caption{An enumerated two dimensional structure  (here for simplicity represented as a rectangle) made of $n_c=32$ cells, with a sub-region $A$ made of $n_c-i=12$ cells, and its complementary region made of $i=20$ cells. Breakable cells are represented in grey color, unbreakable ones in white color.
\label{n-cells}}
\end{figure}

What we need to show is that the  probability
\begin{eqnarray}
P(A|n_c) = {1\over 2^{n_c}-1}\sum_{\rho_{n_c}} P(A|\rho_{n_c}),
\end{eqnarray}
that a cell in region $A$ breaks, when a $n_c$-cell breakable membrane is chosen at random, in a uniform way, is equal to $P(A|\rho_{u;n_c})$. For this, all we have to do is to reorganize the $n_c$ cells forming the hypermembrane on a line, in the following way: we choose a method to enumerate the $i$ cells contained in the complementary region of $A$, place them in order on the left side of a line, then, to follow, we do the same with the remaining $n_c-i$ cells contained in $A$, placing them on the right side (see Fig.~\ref{n-cells}, for a two-dimensional example).
In this way, we transform the multidimensional $\rho_{n_c}$-hypermembrane, made of $n_c$ cells, into an effective 1-dimensional one, made of a same number of cells, and according to (\ref{equality-i}) we have: 
\begin{eqnarray}
\label{universal=uniform}
P(A|n_c) = {1\over 2^{n_c}-1}\sum_{(x\cdots)} P(i\to 0|x\cdots) = {n_c-i\over n_c}.
\end{eqnarray}

\vspace{-0.4cm}
\section{Robustness}
\label{Robustness}
\vspace{-0.4cm}

In Sec.~\ref{theorem} we have proved a theorem  establishing the (probabilistic) equivalence between universal measurements and uniform measurements, and in the previous section we have shown that universal measurements do correspond to an average over measurements describing very different probabilistic models, not necessarily amenable to the Kolmogorovian or Hilbert ones. Also, as emphasized in Sec.~\ref{Therhomodel}, and in the first part of this article \citep{AertsSassolideBianchi2014},  uniform measurements, and therefore also universal measurements, are  equivalent to quantum measurements, whenever the structure of the set of states is Hilbertian. In this section, we investigate another remarkable characteristic of universal measurements: their \emph{statistical robustness}. 

In \citet{AertsSassolideBianchi2014}, we have explained that universal/uniform measurements can be interpreted as processes lying in between the pure discovery, classical regime, and the pure creation, solipsistic\footnote{In \citet{AertsSassolideBianchi2014}, we have used the term `solipsistic' to denote measurements whose unpredictable outcomes are in no way affected by a change of the state of the measured entity.} regime. Another way of characterizing them, is to say that they correspond to a balance between two different kinds of randomness, which are fused together in an experiment. The first one is associated with the lack of knowledge of the experimenter about the nature of the experiment which is actually performed, at each measure (the choice of a specific $\rho$-hypermembrane in the GTR-model), and the second one is associated with the experiment itself, i.e., with the inherent randomness in the process of actualization of a specific deterministic interaction (the \mbox{\boldmath$\lambda$} variable in the GTR-model). 

Interestingly, the difference between these two levels of randomness, or of lack of knowledge, which are distinguishable in theoretical terms, but fused together, and therefore undistinguishable, in practical terms, within the statistics of outcomes of a measurement, is also mentioned in our ordinary language in the distinction between the term \emph{random} and the term \emph{arbitrary}. In some languages, the literal meaning of random  is ``that which falls towards you'' (as for instance a danger, a risk, like when dice are rolled and you may loose a bet). For example in Dutch it is the word `toevallig', from `fallen', which means `to fall', and `toe', which means `towards you'. The same root of meaning is strongly present in the French word `hazard', which indeed means `danger, risk', and the English `hazard'. And arbitrary,  (arbitraire, in French, willekeurig, in Dutch) is ``that which is guided by your will'' (as for instance when you take a decision, for example to perform an experiment, in a given moment). Hence our ancestors, in the era when language developed, were aware of the difference between the randomness provoked by themselves, as subjects, expressed in the word `arbitrary', and the randomness coming from the objects of their experience, expressed in the word `hazard'.  By means of the specific notion of universal measurement, this distinction can be expressed in precise mathematical terms, as the difference between the randomness coming from the level of the choice of the measurement (the selection of a $\rho$), and the randomness coming from the level of execution of the measurement (the selection of a \mbox{\boldmath$\lambda$}). 

This remark allows us to introduce the next analysis we want to present in this article. Although an experimenter has clearly no absolute means to control the amount of randomness which is inherent in a measurement, once it has been selected, s/he may nevertheless have the possibility to exert some control over the available measurements, that is, over the probability densities $\rho$ which can be selected, every time a measurement is performed. In a cognitive experiment with human subjects, such a control might consist of only selecting subjects with certain characteristics 
as participants in the experiment, e.g. individuals with a particular affinity or lack of affinity with certain fields of human experience, belonging to certain schools of thought, having or not having a specific culture, etc.

To account in a simple way for this ability of the experimenter to vary her/his level of control over a (universal) measurement, in the general case of $N$ possible outcomes,  we introduce a ``control region'' $C^\epsilon$ of $S_{N-1}$, of Lebesgue measure $\mu_L(C^\epsilon) = (1-\epsilon) {\sqrt{N}\over (N-1)!}$, $\epsilon\in [0,1]$, such that the only allowed probability densities are those of the truncated form:
\begin{equation}
\rho^\epsilon({\bf z}) = {1\over 1-\int_{C^\epsilon} \rho({\bf z})d{\bf z}}\left\{ \begin{array}{ll}
         \rho({\bf z}) & \mbox{if ${\bf z}\in S_{N-1}\setminus C^\epsilon$}\\
        0 & \mbox{if ${\bf z}\in C^\epsilon$}.\end{array} \right.
\end{equation}
In the physical realization of the GTR-model, one can imagine that the experimenter can apply a special substance on the hypersurface corresponding to region $C^\epsilon$, so that the hypermembranes will become unbreakable in that region, and all  hidden deterministic interactions associated with the  \mbox{\boldmath$\lambda$} belonging to $C^\epsilon$ will become unavailable (i.e., it will not be possible to actualize them during a measurement). 

If we consider $\rho^\epsilon$ of this ``controlled'' kind, the average (\ref{averagerho2}) becomes: 
\begin{eqnarray}
\label{averagerho2-epsilon}
 P({\bf x}\to \hat{\bf{x}}_{i}|n_c;\epsilon)=  {1\over 2^{n_c}-1}  \sum_{\rho^\epsilon_{n_c}}P({\bf x}\to \hat{\bf{x}}_{i})|\rho^\epsilon_{n_c})=P({\bf x}\to \hat{\bf{x}}_{i})|\rho^\epsilon_{u;n_c}),
\end{eqnarray}
where the sum runs  over all  possible cellular probability densities $\rho^\epsilon_{u;n_c}$ that are identically zero in $C^\epsilon$. Thus: 
\begin{eqnarray}
\label{averagerho-limit-epsilon}
\lefteqn{P^{\rm{univ}}({\bf x}\to \hat{\bf{x}}_{i}|\epsilon)\equiv\lim_{n_c\to\infty} P({\bf x}\to \hat{\bf{x}}_{i}|n_c;\epsilon)}\nonumber\\
&& = \lim_{n_c\to\infty} P({\bf x}\to \hat{\bf{x}}_{i})|\rho^\epsilon_{u;n_c}) = P({\bf x}\to \hat{\bf{x}}_{i}|\rho^\epsilon_u),
\end{eqnarray}
where 
\begin{equation}
\rho^\epsilon_u({\bf z}) = {1\over \epsilon}\left\{ \begin{array}{ll}
          {1\over {\sqrt{N}\over (N-1)!}} & \mbox{if ${\bf z}\in S_{N-1}\setminus C^\epsilon$}\\
        0 & \mbox{if ${\bf z}\in C^\epsilon$}\end{array} \right.
\end{equation}
is the truncated uniform probability density, describing an elastic structure uniformly breakable in $S_{N-1}\setminus C^\epsilon$,  and uniformly unbreakable in $C^\epsilon$. Therefore, we have: 
\begin{eqnarray}
P^{\rm{univ}}({\bf x}\to \hat{\bf{x}}_{i}|\epsilon)=\int_{A_i} \rho^\epsilon_u({\bf z}) d{\bf z}
={1\over \epsilon  {\sqrt{N}\over (N-1)!}} \int_{A_i} \chi^{\epsilon}({\bf z}) d{\bf z},
\end{eqnarray}
where $\chi^{\epsilon}({\bf z})$ is the characteristic function of $S_N\setminus C^\epsilon$.

To study the robustness of the $\epsilon$-universal measurement, we consider a state ${\bf x}' = {\bf x} + \delta {\bf x}$, very close to ${\bf x}$, and want to compare the transition probabilities of these two states:
\begin{eqnarray}
|P^{\rm{univ}}({\bf x}'\to \hat{\bf{x}}_{i}|\epsilon) - P^{\rm{univ}}({\bf x}\to \hat{\bf{x}}_{i}|\epsilon)|={1\over \epsilon  {\sqrt{N}\over (N-1)!}} \left|\left(\int_{{A'}_i} -\int_{A_i}\right)\chi^{\epsilon}({\bf z}) d{\bf z}\,\right|,
\label{difference}
\end{eqnarray}
where  ${A'}_i$ is the convex region associated with state ${\bf x}'$. As we said, we assume that the experimenter is able to vary its level of control over the measurement, by varying the parameter $\epsilon$, i.e., by varying the size of the unbreakable region $C^\epsilon$. We also assume that, by adjusting the control parameter $\epsilon$, the experimenter tries to maximize the robustness of the measurement, that is, to minimize the variation of the probabilities with respect to a small variation $\delta {\bf x}$ of the state.  

A  way to do this is to let $\epsilon\to 0$. Depending on how $C^\epsilon\to S_{N-1}$, this will give rise  to either classical (almost) deterministic processes, or to solipsistic-like processes. For instance, if $C^\epsilon = S_{N-1}\setminus B^\epsilon(\mbox{\boldmath$\lambda$})$, with $B^\epsilon(\mbox{\boldmath$\lambda$})$  a $(N-1)$-ball  of volume $\epsilon {\sqrt{N}\over (N-1)!}$ contained in $S_{N-1}$, centered in \mbox{\boldmath$\lambda$}, then $\rho^\epsilon({\bf y}) \to \delta({\bf y} - \mbox{\boldmath$\lambda$})$, as $\epsilon\to 0$, which corresponds to a purely deterministic process. Probabilities will then either be $0$ or $1$, and will typically remain constant when the state ${\bf x}$  is slightly varied (the only possible variations being of course when ${\bf x}$ crosses the boundaries of the region ${A}_{i}$  containing \mbox{\boldmath$\lambda$}, causing the probabilities to abruptly shift from $0$ to $1$, or vice versa, passing through a point of unstable equilibrium). 

Similarly, considering a more general situation where, say, $C^\epsilon =  S_{N-1}\setminus \cup_{i=1}^nB^\epsilon(\mbox{\boldmath$\lambda$}_i)$, with the $B^\epsilon(\mbox{\boldmath$\lambda$}_i)$ that are now balls of volume  ${\epsilon\over n} {\sqrt{N}\over (N-1)!}$,  we have in this case  $\rho^\epsilon({\bf y}) \to {1\over n}\sum_{i=1}^n \delta({\bf y} - \mbox{\boldmath$\lambda$}_i)$, as $\epsilon\to 0$, which corresponds to a mixed condition  describing, for certain states, a deterministic process, and for others a solipsistic one. But also in this case probabilities  will  remain  constant, for small variations of almost all states ${\bf x}$.

So, we can say that the regime $\epsilon\to 0$ exhibits robustness in a trivial way, considering that probabilities become constant in this limit. There is however a more interesting regime, exhibiting robustness in a non trivial way. Indeed, observing that the parameter $\epsilon$ is at the denominator of the  fraction in (\ref{difference}), it is clear that if we increase $\epsilon$, the ratio will decrease. Regarding the second factor in (\ref{difference}), considering that the Lebesgue measure of $C^\epsilon$ tends to 0 as $\epsilon$ tends to 1, it is natural to assume that there exist a $\tilde\epsilon<1$ such that, for $\epsilon\geq \tilde\epsilon$, a same portion of $C^\epsilon$ will be contained in ${A'}_{i}$ and ${A}_{i}$. Accordingly, the difference of the two integrals in (\ref{difference}) will become independent of $\epsilon$, for $\epsilon\geq \tilde\epsilon$, and be equal to $\mu_L( {A'}_{i})-\mu_L( {A}_{i})$, so that for $\epsilon\geq \tilde\epsilon$, we obtain: 
\begin{eqnarray}
|P^{\rm{univ}}({\bf x}'\to \hat{\bf{x}}_{i}|\epsilon) - P^{\rm{univ}}({\bf x}\to \hat{\bf{x}}_{i}|\epsilon)|={1\over \epsilon} \, |{x'}_i - {x}_i| ={1\over \epsilon}\,|\delta x_i|,
\label{differencebis}
\end{eqnarray}
and clearly the most robust condition (i.e., the smoothest possible variation of the probabilities) is obtained by letting $\epsilon\to 1$. 

In other terms, in accordance with the recent 
%Massimiliano: (Bayesian) The author of the article told me that his analysis wasn't at all Bayesian, and asked me to eliminate that mention
analysis of \citet{DeRaedtetal2013}, we find that universal measurements, like quantum measurements, correspond to measurements for which the frequencies of the observed outcomes are maximally robust with respect to small variations of the state of the system, i.e., with respect to small variations in the conditions under which the measurements are carried out. This allows us to  further characterize universal measurements as measurements in which the experimenter exerts the least possible control, letting all the randomness naturally present in  the experimental context freely manifest; or, to put it in an equivalent way, they correspond to measurements  dealing with that ``immanent'' randomness which remains after all forms of control have been subtracted, i.e., all attempts to decrease randomness have been subtracted, by whatever means.

This is typically what happens when dealing with microscopic entities, since we are not usually in a position to control  what happens at the level of the hidden interactions, when performing measurements on these entities, and therefore reduce the  randomness contained in them. Of course, this doesn't mean that such randomness will necessarily always be an unavoidable ingredient of every quantum measurement. Indeed, one can think that, by means of more sophisticated experimental protocols, it may become possible in 
the future to push away at least part of it. Just to give an example, so-called minimally-disturbing implementations of von-Neumann measurements, using  the formalism of positive operator valued measures \citep{Barnum2002}, could constitute a possibility in that direction.

Of course, the same holds true in measurements with human subjects. Indeed, it is in principle always possible to exploit a deeper understanding of the working of the human mind to design protocols where the questions addressed to the subjects become less interrogative, and  more determinative. Mind control techniques used to favor certain reactions from a person, for instance in the ambit of selling, precisely do that. Think how an experienced seller can monitor the customer's face and manner, in real time, so as to find the exact moment to place a certain suggestion, to have a higher probability of producing a certain response. 

It is important however to emphasize that when an experimenter increases her/his level of control in a certain experiment, be it a psychological experiment or a physical one, in a way or another s/he will have to alter the experimental protocol. And by doing so, s/he may end up affecting the nature of the observed properties, as properties are operational quantities, that is, quantities defined by means of specific operations specifying how  they have to be tested (i.e., observed), and if we change these operations (for instance to reduce randomness) we may  end up changing the very (operational) definition of the properties under observation (for a discussion of this point, see \citet{SassolideBianchi2014}). 

In addition to that, when we try to acquire more control, the risk is to also obtain measurements containing lesser information about the state of the measured entity in comparison to that of the state of the measuring system. Indeed, as shown in the Bayesian-like analysis of \citet{AertsS2006}, pure quantum measurements can also be characterized as  experiments in which the observer (i.e., the measuring system) actively attempts to minimize her/his  influence on the produced outcomes. This of course is fully compatible with our characterization of quantum measurements as universal measurements, in which the observer is precisely in a condition of maximal lack of knowledge about the measurement taking place. Indeed, maximal lack of knowledge also means maximal lack of control, and maximal lack of control means minimal influence over the observed entity.

\vspace{-0.4cm}
\section{Conclusion}
\label{Conclusion}
\vspace{-0.4cm}

To conclude, in this second part of the article we have continued our analysis of the GTR-model, emphasizing its broad structural versatility in the description of measurements which are not necessarily characterized by a Hilbertian probability model. We have then used the model to rigorously prove the equivalence between universal measurements and measurements characterized by uniform probability densities $\rho_u$ (the UTR-model), which in turn are compatible with the predictions of the Born rule, whenever the states of the entity under investigation comes from a Hilbert space. 

To define the universal average characterizing universal measurements, we have followed the traditional way of dealing with high levels of randomness present in nature, like in the analysis of Brownian motions, considering first a discretization of the problem. More precisely, by approximating general $\rho$-hypermembranes as limit of finite cellular structures, we have succeeded in defining in a mathematically precise and physically transparent way the fundamental notion of \emph{universal measurement}, describing situations where the lack of knowledge is double, i.e., not only about the specific (almost deterministic) hidden interactions which are actualized during an experiment, but also about how such interactions are each time selected.

As an additional element of characterization of universal measurements, we have also modelized the situation in which an experimenter decides to take some active control over the randomness present in a given measurement context, by making certain hidden interactions unavailable. This active control (which corresponds to an acquisition of knowledge) produces experiments whose statistics are less robust with respect to small variations of the initial conditions, which  explains why experimenters will have the tendency to increase the universality of their measurements, i.e., to increase the reach of the averages that are possibly subtended by their measurements. This means that a condition of lowest possible control, i.e., lowest possible knowledge regarding the nature of the measurement which is each time conducted, is the most favorable one in terms of the reproducibility of the obtained statistics of outcomes.

We conclude by observing that the notion of universal measurement opens to different possibilities regarding the way we should interpret and collect data in future cognitive studies, and it is the plan of the present authors to elaborate further this notion, both theoretically and experimentally, with the aim of increasing further our understanding of the fundamental probabilistic structure underlying human cognition and decision.

\end{document}